# Multi-sphere approximation of realistic particles with a combined 3D thinning and greedy algorithm

Fei-Liang Yuan (FeiLiang.Yuan@gmail.com)[1]

[1]*Multiphase Flow Systems, Institute of Process Engineering, Otto-von-Guericke-University Magdeburg, Hoher Weg 7b, D-06120, Halle (Saale), Germany*

## Abstract

This article is the updated version of a previously published paper: "Combined 3D thinning and greedy algorithm to approximate realistic particles with corrected mechanical properties, Granular matter (2019)" by the author [58]. The main changes here are: (1) a recently developed DEM code Rigid3D is used for the DEM simulations, instead of the open-source software LIGGGHTS in the original work; (2) three new validation tests were carried out.

The main idea of this work is to combine the 3D thinning and greedy set-covering algorithms for the approximation of realistic particles with multiple spheres. First, the particle's medial surface (or surface skeleton), from which all candidate (maximal inscribed) spheres can be generated, is computed by the topological 3D thinning algorithm. Then, the clump generation procedure is converted into a greedy set-covering problem. The advantage of this approach over previous studies is that, with a given particle media surface and a grid resolution for the discretized particle, a minimum number of component (or primary) spheres for an MS particle is guaranteed for a maximum volume coverage, thanks to the combined algorithm. To correct the mass distribution of an MS particle due to the highly overlapped component spheres, linear programming (LP) is used to adjust the density of each component sphere, such that the aggregate properties such as mass, center of mass and inertia tensor are identical or close enough to the prototypical particle. We then demonstrate the capability of the combined algorithm to approximate simple to complex particle shapes in three test cases: (1) particle-wall contact; (2) particle settling; and (3) granular flow in a rotating drum.

---

## 1. Introduction

### 1.1 Background

Granular materials are often encountered in many natural and industrial processes. Typical examples include particle transportation within fluid such as sedimentation and erosion in waterways or coastal areas, aeolian sand movement in deserts, and airborne particles in indoor environments; conveying, mixing and separation of dry or wet granular materials in chemical, mineral, pharmaceutical and food processing [8, 9, 32, 52], and the list goes on. To study the fundamental mechanisms or obtain optimized design parameters in such processes, scaled or pseudo two-dimensional (2D) laboratory experiments are critical to obtain information that characterizes granular systems such as angle of repose, velocity and local porosity profiles. However, experimental results and derived empirical formulas from scaled laboratory models may lead to considerable deviation when extrapolated or applied onto prototype values due to dissatisfaction of similarity laws, while a laboratory model with prototypical size is sometimes extremely expensive or even impossible. Moreover, for a fully three-dimensional (3D) granular system, it is difficult to utilize high-speed digital image analysis technique to capture the internal motion of particles due to their opaque nature, while invasive probes may disturb the system (at least locally) and hence affect the experimental results [14].

As an important alternative for physical experiments, numerical simulation technique, such as discrete element method (DEM) introduced by Cundall and Strack [13], does not suffer from these problems, even though it does suffer numerically inherited effects such as computational cost, truncation error due to the second-order velocity-Verlet integration scheme [25] and inadequate accuracy of particle shape approximation, etc. In the soft-sphere framework used in DEM, the position and velocity of individual particle are updated using Newton's second law at each explicit time-step. The resulting force acting on an individual particle is calculated by summing up the contact forces due to the particle-particle, particle-wall and particle-fluid interactions, and non-contact forces due to gravity, electric and magnetic fields. In this sense, the particle-scale information can be obtained from DEM simulations, which is essential for understanding of the complex dynamics of granular systems.

Particles encountered in nature and industry are mostly irregular-shaped, and the effect of particle shape has a strong impact on the particle-scale and macro-scale behaviour of granular systems [29, 49, 57], therefore the realistic particle shape has to be modelled properly in DEM simulations rather than using a simple sphere. Generally there are two approaches to model a 3D particle with a realistic shape: multi-sphere (overlapping or not) and polyhedron or smoothed-polyhedron (Minkowski sum of a polyhedron with a sphere, also called sphero-polyhedron), because ellipsoid, super-quadric [7, 39, 47] or any other mathematically described shapes usually have symmetrical, continuous and smooth surfaces, thus are not sufficient to represent realistic particles that are usually asymmetrical and angular.

In recent years, polyhedral and smoothed-polyhedral approximation of particle shapes have received a broad attention in DEM community, due to their straightforward and versatile description of realistic particles which allows either sharp or rounded vertices/edges [29, 54]. One of the main challenges in DEM simulations using polyhedrons is the accurate contact detection and resolution between a pair of particles. The collision handling of multiple particles is usually performed in two independent phases: the broad phase and the narrow phase. The purpose of the broad phase processing is to quickly find a list of neighbour particles that are potentially colliding with a given particle, such approaches include spatial partitioning and bounding volumes [17]. The narrow phase is responsible for the actual collision detection, and calculation of contact forces between two potentially colliding particles, once the neighbour list for each particle is established in the broad phase processing.

A brief overview of generic contact algorithms for the narrow phase processing is given in the following. The *Common Plane* (CP) algorithm [12, 44] simplifies complex polyhedron-polyhedron intersection into polyhedron-plane contact problem, once the CP between two polyhedral particles is established. Nevertheless, the identification of actual contact points on the CP is still computationally intensive according to Höhner et al. [27]. Since convex polyhedron can be described as the intersection of a set of half spaces, the contact detection problem can be done mathematically, i.e., computing the intersection of two sets of *Half Spaces* [36, 43], if the result is empty, then the two particles are not in contact, otherwise the intersection of two sets of half spaces (also a convex polyhedron) is the exact overlap volume that can be used to calculate the contact force. The *GJK* (Gilbert-Johnson-Keerthi) algorithm is elegant and fast to calculate the overlap distance and contact point between two sphero-polyhedral particles (sweeping a small sphere around the profile of polyhedron), based on the concepts of support function and Minkowski difference [2, 22, 42, 53]. Depending on the sweeping radius, the edges of polyhedral particle can be nearly-sharp or rounded. Moreover, for support functions work with shapes such as cylinder, sphere, ellipsoid or even super-quadric, the actual particle shape is not necessarily modelled by very fine mesh, which can lead to poor performance for those methods only applicable to ideal polyhedrons (e.g. Common Plane and Half Space based algorithms). Dong et al. [15] proposed a general approach to calculate contact forces based on the *pre-calculated overlap information*. By discretizing the particle body into small cells in 3D space, the overlap volume is simply the intersection of two cell sets that present two colliding particles, which shares some similar feature with the Half Space algorithm. Although the calculation of overlap information is one-off cost, this method might not be feasible for those systems with many particle shapes and sizes, as the computational cost and memory usage to build/store the database might not be even affordable for a computer cluster.

It is important to note that the aforementioned algorithms only work with convex shapes, while the realistic particles are either convex or non-convex (concave), and the rolling resistance and interlocking between non-convex particles tend to be larger than that of packing of convex particles, due to the increased angularity [38, 41]. In fact, non-convex polyhedron can be decomposed into multiple convex components [21] and treated as a composite shape of convex sub parts. This technique is commonly used in several physics engines [e.g. 11] for robotics and game development. Nevertheless, according to the recent review [40] on the development of non-spherical granular systems, there is little progress in the modelling 3D non-convex particles using polyhedrons in DEM codes. Until very recently, Rakotonirina et al. [48] proposed a *glued-convex method*, which is similar to the composite shape concept, to deal with concave particles using GJK algorithm. While Kawamoto et al. [31] used *Level Set* functions to describe particle shapes, and node-to-surface contact algorithm to handle convex and non-convex contact problem, it seems that the method is more difficult to implement than the composite convex method using GJK algorithm, in terms of shape description and contact resolution.

As mentioned earlier, the collision detection and contact forces calculation between convex polyhedral particles are complex, let alone the non-convex particles. Using multi-sphere method [1, 18, 20, 37, 55], i.e., approximating an arbitrarily shaped particle (convex or non-convex) by a set of "glued" spheres, can convert complex particle-particle interaction into simple sphere-sphere contact problem. For clarity, a particle made out of multiple spheres is termed as "clump" in this work. In the multi-sphere method, positions of clump's spheres are fixed in the particle body frame, and sphere-sphere interactions inside the clump are ignored by not including them in the neighbour list. The clump's translational and angular velocity is updated using the resultant contact forces, body forces and torques (about the particle's center of mass) acting on all component spheres. The contact between clump-wall, clump-clump are simply handled by all sphere-wall, sphere-sphere contact pairs.

One issue associated with the multi-sphere approach is the occurrence of multiple contact points [40], as the surface of clump is the boolean union of all component spheres which usually yields a rather bumpy surface. Imagine a perfect rigid sphere impacting a flat wall, there will be only one contact point, whereas there can be more than one contact point if this sphere is approximated by multi-sphere approach. As a result, the effect of multiple contact points on the normal and tangential forces evolution may lead to considerable deviation between multi-sphere approach and accurate solution as shown in [26, 33, 35]. To overcome this issue, Höhner et al. [26] proposed an incremental approach to calculate contact forces. The main idea is to divide the incremental contact forces during each time-step by the number of active contact points, in order to avoid the accumulation of numerical error. By adding the averaged increments to the forces calculated from previous DEM time-step, we get the current step values. Höhner et al. [26] also compared the deviation of normal and tangential forces of single spherical and ellipsoidal particles impacting a flat wall with reference solution. The results showed that the deviation was significantly reduced by using the incremental approach, and surprisingly the average deviation of multi-sphere approach is smaller that the polyhedral approach, in the case of ellipsoid-wall collision where the approximation accuracy is 15-200 spheres and 15-200 vertices, respectively.

Despite the effect of multiple contact points on the single particle level, this effect may compensate each other among colliding particles in large granular systems [35]. Moreover, the artificial roughness introduced by the multi-sphere approach is not necessarily a drawback for highly irregular-shaped particles. Höhner et al. [29] carried out experiments and numerical simulations on the granular media flow in a rectangular hopper. Five different particle shapes including sphere, icosahedron, dodecahedron, hexahedron and wood cylinder were approximated using multi-sphere, super-ellipsoid, polyhedral and smoothed polyhedral approaches. The simulations results showed a good agreement with the experiments, and only a minor difference has been found between the multi-sphere and other approximation approaches. In this sense, the multi-sphere approach is still worth studying due to its simplicity on handling the interaction of convex and non-convex particles, despite the recent progress made in the polyhedral approaches.

## 1.2 Related work

Approximating a realistic particle with multiple spheres can be highly non-trivial, as the aim of mutli-sphere approach is to fill a particle's body (e.g. surface mesh from 3D scanner) tightly with minimum spheres, while keeping the shape approximation at an acceptable accuracy level. Component spheres of a clump may overlap each other or not. In the case of non-overlapping, thousands of spheres are usually required for a good approximation [55, 56], thus prohibits its application for a system with large amount of particles, while overlapping-sphere representation tends to produce smoother surface with less spheres. At a broad level, 3D overlapping-sphere algorithms presented in previous studies might be classified into several catalogues in terms of the pre-processing to generate candidate spheres, and briefly over-viewed as follows.

***Medial surface***: The medial surface, which corresponds the surface skeleton of a 3D object, is simply the locus of the centers of all maximal inscribed spheres that have at least two closest points on the object's boundary [5]. Since constructing an exact medial surface for an irregular-shaped object is complex and computationally expensive, Hubbard [30] proposed a fast approach to approximate the medial surface based on the Voronoi diagram formed by the points on the object's surface. The generated Voronoi vertices inside the object represent points roughly lying on the medial surface. Therefore each candidate sphere can be defined with a Voronoi vertex as the center, and radius being the distance from Voronoi vertex to its forming points. The remaining task is reducing the number of spheres while preserving as much approximation accuracy as possible. The sphere-reduction strategies such as merging adjacent spheres, bursting one of the spheres and using surrounding spheres to fill the gap, aim to generate new sub-set of spheres that still cover the object surface. For more information on these reduction algorithms one might refer to the literature [6, 30].

***Surface points***: Rather than constructing medial surface, candidate spheres (i.e., maximal inscribed spheres or medial-spheres) may be directly approximated from the surface points of 3D objects. Ferellec and McDowell [18] proposed a straightforward method to obtain such spheres: from a random surface point on the 3D object and its inward point normal, a sphere can be generated whose radius vector is from a point on the normal to the surface point. The radius (initially zero) is increased gradually until the sphere surface touches another surface point on the object. At this point, the sphere is considered as maximal inscribed. Another similar approach presented in Taghavi [50] uses 3D object's Delaunay tetrahedral mesh to obtain those candidate spheres that are simply the circumscribed spheres for each tetrahedron. It should be noted that the generated spheres from both methods do intersect with the object's surface mesh. Therefore, in order to reduce the error, the points on the object's surface mesh must be dense enough, such that the gap (outside of the object) between inscribed spheres and the object's surface is of several orders of magnitude smaller than the object size (e.g. equivalent diameter). In addition, tuning parameters that control the sequential sphere-inserting process have significant impact on the surface smoothness and number of spheres per clump, thus bring some uncertainty to the final result.

***Uniform gird***: In the uniform grid based methods [20, 37], the 3D object is first discretized into numerous small voxels (i.e., unit cubes). For each voxel a candidate sphere can be generated such that the sphere is centered on the voxel and tangential to the inner surface of the 3D object. Once all candidate spheres (equal to the number of voxels) are computed and sorted into descending order of radius, greedy algorithm is then used to sequentially insert a single sphere that has maximum coverage of voxels (excluding voxels already covered by previously inserted spheres), either with the constraint that newly inserted sphere must be connected to previously inserted spheres to ensure the clump is continuous [20], or without any constraint [37]. Uniform gird based methods are easy to implement since we don't have to compute all the candidate spheres roughly located on the medial surface of 3D objects; however, the drawback is that a large number of candidate spheres is inevitable if we need more accurate (fine voxel) shape approximation, as a result the sphere-inserting process is more computationally expensive (several order of magnitude) than that of coarse version.

Another noteworthy work by Phillips et al. [46] presented a medial axis and heuristic based approach for optimal filling of arbitrarily shaped polygons with disks. Also, some analytical formulations for the spatial distribution of the disks were derived. However, the findings of their work are based on some simplified assumptions in two-dimensional (2D) space, and may not hold on in 3D space (i.e. not applicable for 3D objects). Further development of practical methods for finding optimal solutions in 3D space is needed.

## 1.3 Objective of this study

The aforementioned 3D overlapping-sphere methods to model realistic particles have their pros and cons: the medial-sphere based methods tend to produce less candidate spheres, thus are computationally efficient in the sequential sphere-inserting process; however, generated clump is not guaranteed to be composed of minimum spheres, as the outcome mainly depends on the user-defined input parameters. Whereas in the uniform grid based methods, as the sphere selected at each step has maximal effective coverage (amount of unit cells that fall in the sphere and are not covered by previously selected spheres) thanks to the greedy algorithm. Thus the number of spheres per clump is optimized with a given number of candidate spheres. Nevertheless, the majority of the candidate spheres are redundant as they are far from the underlying medial surface, and a compromise must be made between the computational efficiency and the level of discretization (i.e., size of unit cube).

Base on the investigation above, the main objective of this study is to combine the concept of medial surface with greedy algorithm. First the 3D object (particle) is voxelized same as the uniform grid based methods, then apply the 3D thinning algorithm based on critical kernels [4] to obtain the medial surface made of voxels, on which all candidate spheres can be generated. To further speed up the sphere-inserting process using greedy algorithm, a secondary grid with fine voxels on the particle boundary and coarse voxels inside the particle is used.

The remainder of this work is organized as follows. Section 2 briefly describes how to obtain medial surface of realistic particles using the state-of-the-art 3D thinning technique. Section 3 presents a greedy algorithm based on unstructured/non-uniform grid to find minimum number of spheres to represent a realistic particle. Three validation test cases are carried out in section 4 to verify the accuracy of multi-sphere models, and compared with analytical and experimental results. Contribution of this study and recommendations are summarized in section 5.

## 2. Computation of medial surface

The reason why the topological 3D thinning is chosen over other surface skeletonization techniques such as Voronoi diagram and distance field based methods [51], is that it is uniform grid based (see Figure 1) and easy to implement. In addition, the secondary non-uniform grid for use with a greedy algorithm can be generated during the particle's voxelization process. Furthermore, in the 3D skeletonization methods in which the particle's surface mesh is required to be fine enough, the amount of vertices of computed medial surface is usually close to the number of vertices of the input mesh (which is still considerable). While the resolution of medial surface (i.e., amount of voxels) calculated by the 3D thinning algorithm is controlled by the size of voxel, thus significantly less candidate spheres can be generated but still achieve same or slightly less coverage of the particle's volume than that of large number of candidate spheres.

### 2.1 Notions for 3D thinning

The very first step of the 3D thinning is converting a particle's surface mesh to a 3D binary image, which is illustrated by a 2D schematic of such process in Figure 1. Given a triangulated mesh enclosing a volume, an axis-aligned bounding box (AABB) slightly larger than the input mesh is computed and discretized into numerous voxels. The particle's 3D binary image is then obtained by collecting those voxels whose centers are inside or on the surface boundary of the input mesh as depicted in Figure 1b. In practice, the whole binary image is stored in a flattened 1D array with values of either 1 (gray cells) or 0 (white cells). Usually a voxel size of $\frac{d_{eq}}{100}$ can make a good approximation of the particle's shape, where $d_{eq}$ is the diameter of sphere of equivalent volume (particle).

We will use some basic notions described in the critical kernels based 3D thinning scheme by Bertrand and Couprie [4].

Let $\mathbb{Z}^3$ denote the 3D digital space. A *d-face* where $d \in \{3, 2, 1, 0\}$ is a $d$-dimensional face of $\mathbb{Z}^3$. Here, a 3-face is a unit cube or voxel; 2-face, 1-face and 0-face correspond a square, a line segment

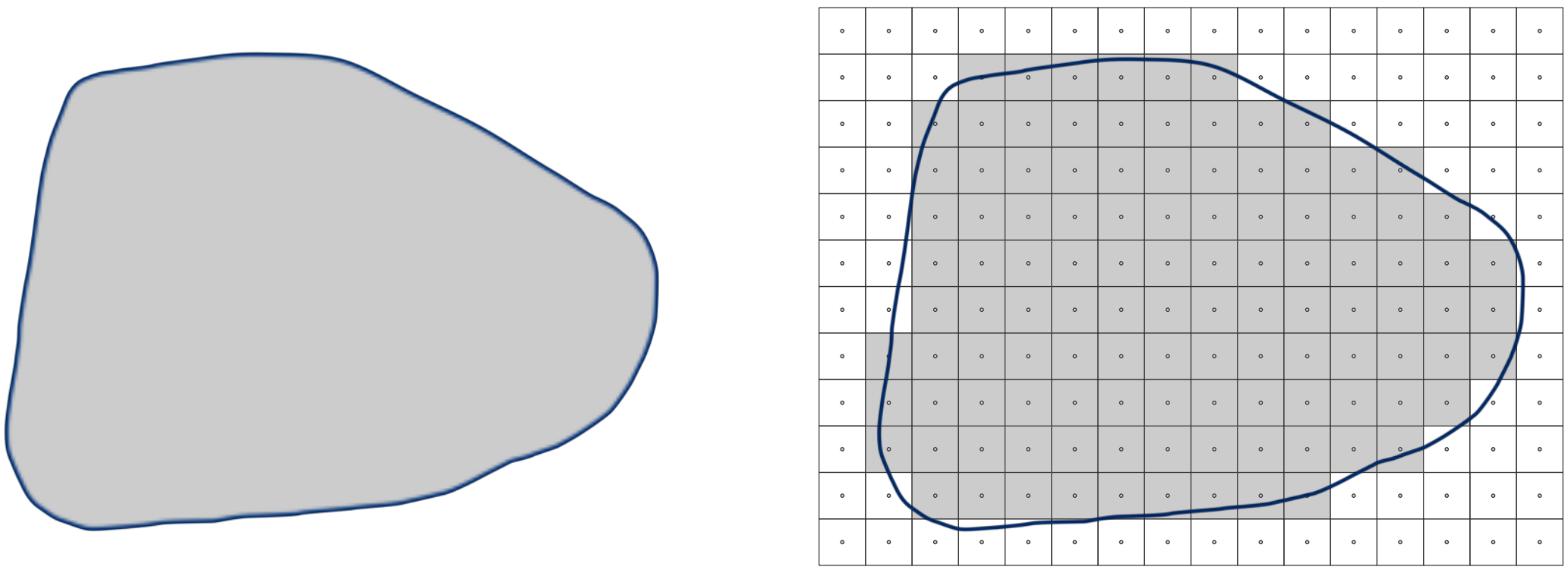

**(a)** Particle **(b)** Voxelized particle - binary image and a point of unit cube, respectively. Let $v_c$, $v_n$ denote a voxel and one of its neighbouring voxel, if $v_c \cap v_n$ is a d-face where $d \in \{2, 1, 0\}$, we say that $v_n$ is a *d-neighbour* of $v_c$, or vice versa. In Figure 2a, we can see that voxel $v_c$ has six 2-neighbours (blue disks), twelve 1-neighbours (green squares) and eight 0-neighbours (orange stars).

**Figure 1:** 2D schematic of particle voxelization.

Let $\mathcal{N}_6(v_c)$ be the set of central voxel $v_c$ that contains $v_c$ and its six 2-neighbours. The set $\mathcal{N}_{18}(v_c)$ contains $\mathcal{N}_6(v_c)$ and twelve 1-neighbours; the set $\mathcal{N}_{26}(v_c)$ contains $\mathcal{N}_{18}(v_c)$ and eight 0-neighbours. $\mathcal{N}_j^*(v_c) = \mathcal{N}_j(v_c) \setminus \{v_c\}$ where $j \in \{6, 18, 26\}$.

One of the core operations in 3D thining is identification of removable voxels. Let $X \in \mathbb{Z}^3$ denote the set of voxels whose value are 1, i.e., the 3D binary image of voxelized particle, $\overline{X}$ the set of white (background) voxels whose values are 0. A voxel $x \in X$ is said to be removable or *simple*, if its removal from $X$ "does not change the topology of $X$".

Let $S$ be a subset of $X$, $S$ is said to be *d-connected* where $d \in \{2, 1, 0\}$, if any two voxels in $S$ can be connected by a path, i.e., a 3D curve made of voxels. In this 3D curve, the intersection of any two adjacent voxels is at least a $d$-face where $d \in [d, 2]$ $(d \leq 2)$. Therefore, a simple voxel can be identified by the connectedness of its neighbourhood configuration.

**Theorem 1** [3] A voxel $x \in X$ is *simple* if and only if:
(i) The set $\mathcal{N}_{26}^*(x) \cap X$ is not empty and 0-connected; and
(ii) The set $\mathcal{N}_6^*(x) \cap \overline{X}$ is not empty and 2-connected in $\mathcal{N}_{18}^*(x) \cap \overline{X}$.

The concept of simple voxel can be extended to simple clique [4]. A clique is a set of mutually adjacent voxels. A voxel set $C \in X$ is said to be a *d-clique* where $d \in \{3, 2, 1, 0\}$, if $\cap\{x \in C\}$ is a $d$-face. Here $d$ is the rank of clique $C$. We note that a clique made of only one voxel is a 3-clique, i.e., any single voxel $\in X$ is a 3-clique (Figure 2a). For a given central voxel $v_c \in X$, a 2-clique can be detected if there exists a 2-neighbour of $v_c$ in $X$. In Figure 2b, $\{v_c, v_1\}$ is a 2-clique. Likewise, a 1-clique exists if one of 1-neighbours $(v_1)$ of $v_c$ is in $X$ as shown in Figure 2c. Here $v_c$ and $v_1 \in X$ is mandatory, voxels $v_2, v_3$ marked by ? are either in $X$ or $\overline{X}$. Therefore a 1-clique is defined by $\{v_c, v_1, v_2, v_3\} \cap X$. A 0-clique for $v_c \in X$ can be found if any of its 0-neighbours exists, or there exist three voxels (including $v_c$) that are mutually 1-neighbours as shown in Figure 2d. In either case,

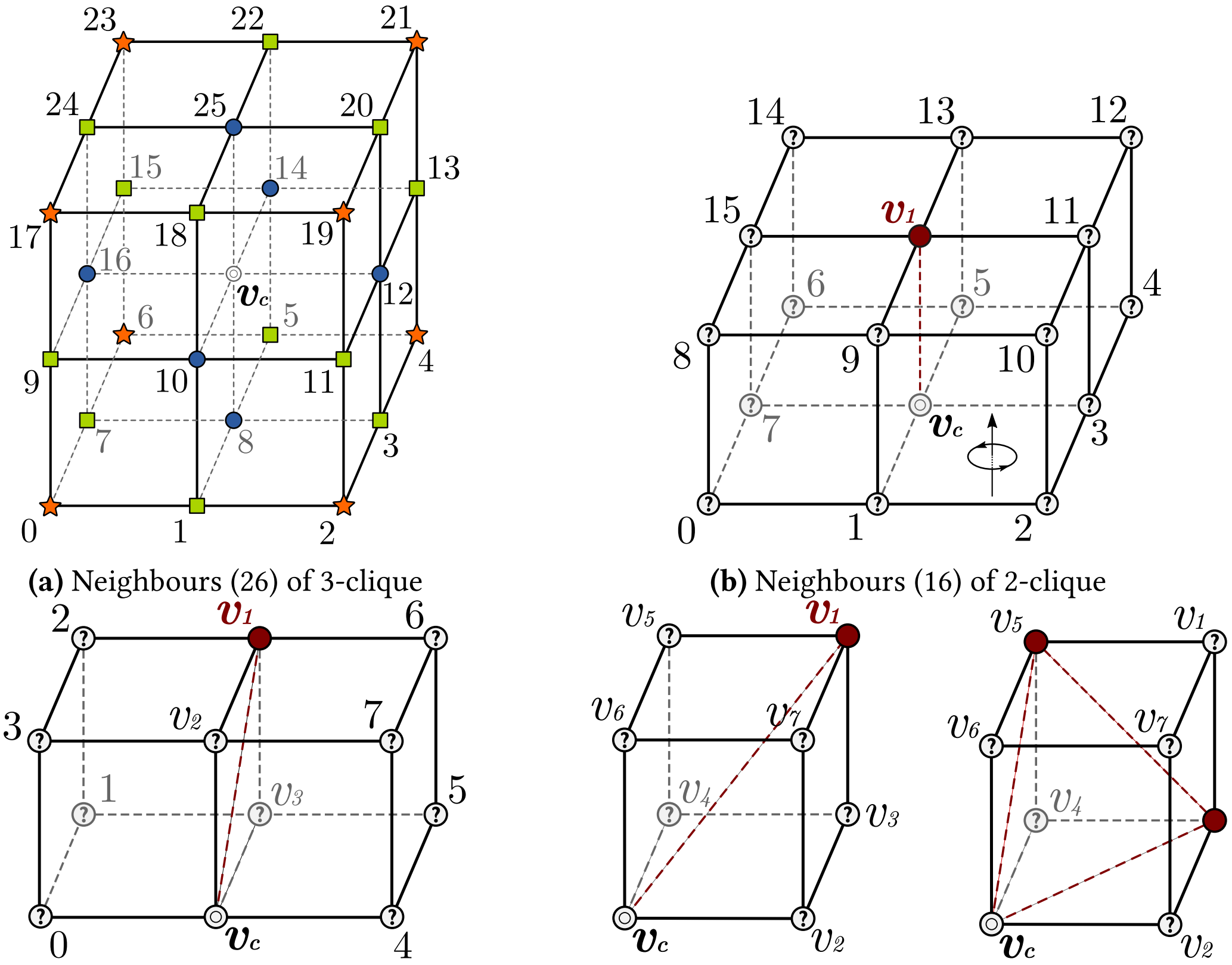

**(a)** Neighbours (26) of 3-clique

**(b)** Neighbours (16) of 2-clique

**(c)** Neighbours (8) of 1-clique

**(d)** 0-clique has no neighbours

**Figure 2:** Neighbourhood and indexing schemes for $d$-cliques where $d \in \{3, 2, 1, 0\}$. Here voxels are represented by their centers. $v_c$ and voxels marked by red disk are in $X$, voxels labelled by question mark within a circle are either in $X$ or $\overline{X}$.

the 0-clique is equal to the set $\{v_c, v_1, v_2, v_3, v_4, v_5, v_6, v_7\} \cap X$.

Let $\mathcal{N}(C_i)$ denote the set of voxels that are adjacent to each voxel in an $i$-clique $C_i$ where $i \in \{3, 2, 1, 0\}$. For $i = 3$, $C_i$ is a single voxel (e.g. $v_c$), thus $\mathcal{N}(C_i)$ is equal to $\mathcal{N}_{26}(v_c)$. For a 2-clique $C_2$, there are 16 voxels that are adjacent to both $v_c$ and $v_1$ as depicted in Figure 2b. A 1-clique has 8 adjacent voxels (Figure 2c), while a 0-clique has no adjacent voxels (Figure 2d). $\mathcal{N}^*(C_i)$ is equal to $\mathcal{N}(C_i) \setminus C_i$.

For an $i$-clique $C_i \in X$, if the set $\mathcal{N}^*(C_i) \cap X$ can be reduced to a single voxel by sequentially removing simple voxels from $\mathcal{K}_i$, we say that $C_i$ is *regular*, which is similar to the term *simple* for voxels. In fact, a 3-clique, i.e., a single voxel is regular, if and only if this voxel is simple. If $C_i$ is not regular, then $C_i$ is *critical*. For example, if a single voxel is critical, it must be non-simple.

## 2.2 Parallel 3D thinning

Despite many notions defined above, the actual 3D thinning process for a voxelized particle is rather straightforward: voxels are removed layer by layer in a topology-preserving way until only the "skeleton" of the particle is left.

**Theorem 2** [4] Let $Y$ be a subset of $X$. If any critical clique in $X$ contains at least one voxel of $Y$, we say that $Y$ is a thinning of $X$.

Originally the notion of critical kernels [4] is based on the *traces* of critical cliques, where the trace of a clique $C_i$ is defined by $\cap\{x \in C_i\}$. Nevertheless, we define the critical kernel of $X$ as the union of all critical cliques in $X$ here for its simplicity. For any subset $Y$ in the critical kernel of $X$, those voxels that are not in $Y$ can be removed in parallel without changing the topology. Thus the smaller the subset $Y$ is, the more voxels can be removed from $X$ at each thinning iteration.

Since a critical clique can have one up to eight voxels, there are many possibilities to construct a thinning subset $Y$ for $X$ that contains at least one voxel of every critical clique. Among all the possible choices, a smaller $Y$ in size (number of voxels) is preferred. Imagine if there exist a critical 3-clique $\{v_2\}$ and a critical 2-clique $\{v_1, v_2\}$ in $X$, if we randomly choose voxel $v_1$ from the 2-clique first, then $v_2$ in the 3-clique must be kept in order to satisfy Theorem 2. However, if we choose a voxel from cliques with higher rank first, in this case, the 3-clique, only $v_2$ is chosen for $Y$ as it is already included in the 2-clique. Therefore, following the decreasing rank strategy [4] (clique rank: $3 \rightarrow 0$), exactly one voxel for each critical clique is kept to ensure that $Y$ has the least number of voxels. Moreover, in order to avoid the uncertainty in selecting, the voxel of lowest array index (recall that all voxels are stored in a 1D array) in a critical clique is taken.

Apart from the voxels that are necessary for preserving the topology of $X$, we have to keep other voxels, so-called *skeletal voxels*, in order to obtain the surface skeleton (i.e., medial surface) or curve skeleton (i.e., medial axis) of $X$. In fact, if we do not keep these skeletal voxels, only one voxel is left after thinning process for those 3D objects without holes and cavities.

Skeletal voxels correspond to the characterized shape features of $X$. If we want to compute surface skeleton, the *surface-end* voxels should be kept at each thinning iteration. A voxel is said to be a surface-end voxel, if it is a *border* voxel, and has no 2-neighbours that are *interior* voxels [45]. A voxel is said to be interior, if and only if it has precisely six 2-neighbours. A voxel is a border voxel if it is not an interior voxel.

Based on the notions of critical clique and surface-end voxel, now we are ready to implement the 3D thinning process for obtaining the particle's surface skeleton made of voxels. The pseudo-code is listed as follows.

To obtain a thinning subset $Y$ for $X$ at each iteration, there are four sub-iterations (line 6-10 in Algorithm 1) following the decreasing clique-rank strategy. First all non-simple voxels (i.e., critical 3-cliques) of $X$ that are not surface-end voxels (line 5) are added into $Y$; then all critical 2-cliques of $X$ that do not include any previously selected voxels (stored in $Y$) are considered for the the function select($C_i$) at line 9, which is responsible for selecting a unique voxel with lowest index from each $i$-clique $C_i$. The same rule applies to critical 1-cliques and 0-cliques, and finally we will get a thinning subset $Y$ that satisfies Theorem 2.

Once $X$ is thinned to $Y$ (line 11), the set of skeletal voxels $K$ needs to be updated (line 12-13), because some border voxels are removed and new surface-end voxels appear. Steps 4-14 are repeated until $X$ can not be thinned further, i.e., at certain iteration the thinning subset $Y$ is equal to $X$. At this point, $X$ is the final surface skeleton of the input mesh.

**Algorithm 1:** 3D surface thinning scheme

**Input:** Triangulated surface mesh of particle: *mesh*
**Output:** Surface skeleton of particle: $X$

**function** surfaceThinning (*mesh*)
  $X$ = voxelize (*mesh*);
  $K = \emptyset$; *// array to store surface-end voxels, initially empty*
  **repeat**
    $Y = K$;
    **for** Rank $i = 3 \rightarrow 0$ **do**
      $T = \emptyset$;
      **foreach** *critical i-clique* $C_i$ *of* $X$*, and* $C_i \in X \setminus Y$ **do**
        $T = T \cup \{\text{select}(C_i)\}$;
      $Y = Y \cup T$;
    $X = Y$; *// $X$ is thinned to $Y$*
    **foreach** *surface-end voxel* $x$ *that is included in* $X \setminus K$ **do**
      $K = K \cup \{x\}$;
  **until** *$X$ can not be thinned further*
  **return** $X$;

## 2.3 Implementation

The key to implement the 3D surface thinning algorithm is the detection of critical cliques [4], as the surface-end voxels can be directly identified by the definition.

To check if a 3-clique (voxel) $v_c \in X$ is regular or simple, we verify if the conditions (i) and (ii) in Theorem 1 are both satisfied by breadth-first search (BFS) algorithm [10]. Since voxel $v_c$ has 26 neighbours whose values are either 1 (in $X$) or 0 (in $\overline{X}$), $\mathcal{N}^*_{26}(v_c)$ has total $2^{26}$ possible configurations. The neighbourhood indexing scheme for $v_c$ is depicted in Figure 2a, thus we can form a "26-bit" positive integer $N_{cfg}$ as the code of each configuration. This code $N_{cfg}$ is calculated by $\sum_{j=0}^{25} 2^j \cdot b(v_j)$, where $b(v_j)$ is the binary value of voxel $v_j$. For each neighbourhood configuration of $v_c$, Theorem 1 is tested, here we denote the result by 0 (simple) or 1 (non-simple). Using the configuration code $N_{cfg}$ as index and the result (0 or 1) as input, the pre-calculated data are stored in a lookup table with $2^{26}$ entries. Therefore, the critical 3-clique detection is converted to a much cheaper array indexing operation, if the lookup table is loaded into memory beforehand.

Likewise, we can build lookup tables for critical 2-cliques and 1-cliques detection. Let $\mathcal{K}_i = \mathcal{N}^*(C_i) \cap X$ where $i \in \{2, 1\}$, if the set $\mathcal{K}_i$ can be reduced to a single voxel by sequentially removing a random simple voxel for $\mathcal{K}_i$, we say that the $i$-clique $C_i$ is regular, otherwise $C_i$ is critical. If we define the orientation of a 2-clique or 1-clique as the vector $v_c \rightarrow v_1$ (voxel center $v_c$ to voxel center $v_1$), the indexing scheme (Figure 2a) for a given central voxel $v_c$ can be mapped to the local neighbourhood indexing schemes for 2-clique and 1-clique as shown in Figure 2b and 2c. Let $N = \mathcal{N}^*_{26}(v_c)$ denote the array to store the neighbours of 3-clique $v_c$. If $N[25] \in X$, a 2-clique $\{v_c, v_1\}$ with $v_1 = N[25]$ is detected, and its 16 neighbours is defined by the set $\{N[9], N[10], N[11] \ldots N[24]\}$.

Similarly, if $N[22] \in X$, a 1-clique $\{v_c, v_1, v_3, v_4\} \cap X$ with $v_1 = N[22], v_2 = N[25], v_3 = N[14]$, and its 8 neighbours $\{N[16], N[15], N[23], N[24], N[12], N[13], N[21], N[20]\}$ is decided. Note that a 0-cliques is necessarily critical as it has no neighbours (i.e, can not be reduced to one voxel).

If we want to compute curve skeleton, *curve-end* voxels in $X$ should be kept. A curve-end voxel $v_c$ is detected if there is only one voxel in the set $\mathcal{N}^*_{26}(v_c) \cap X$. For those 3D objects whose analytical surface skeletons are 3D curves (e.g. sphero-cylinder), curve-ends voxels should be used, because surface skeleton, which is very sensitive to the noise of input mesh, may contain numerous spurious branches.

## 2.4 Verification examples

To verify the implementation, four different shapes from regular to irregular, and their computed surface skeletons, are compared in Figure 3. Because a surface skeleton obtained from the 3D thinning differs over the orientation of the input mesh (slightly at 90 degrees rotation), the principal axes the original particle are aligned to the global coordinate axes, which usually makes the best result in terms of least noise on the surface skeleton.

For simple shapes like box and ellipsoid ($x^2 + \frac{y^2}{2} + z^2 = 1$) as shown in Figure 1a and 3c, the 3D thinning algorithm can produce neat surface skeletons (see Figure 3b and 3d). Note that the surface skeleton of the ellipsoid shown in Figure 3d is computed by keeping the curve-end voxels, as the analytical solution is a line segment (i.e., the major principal axis).

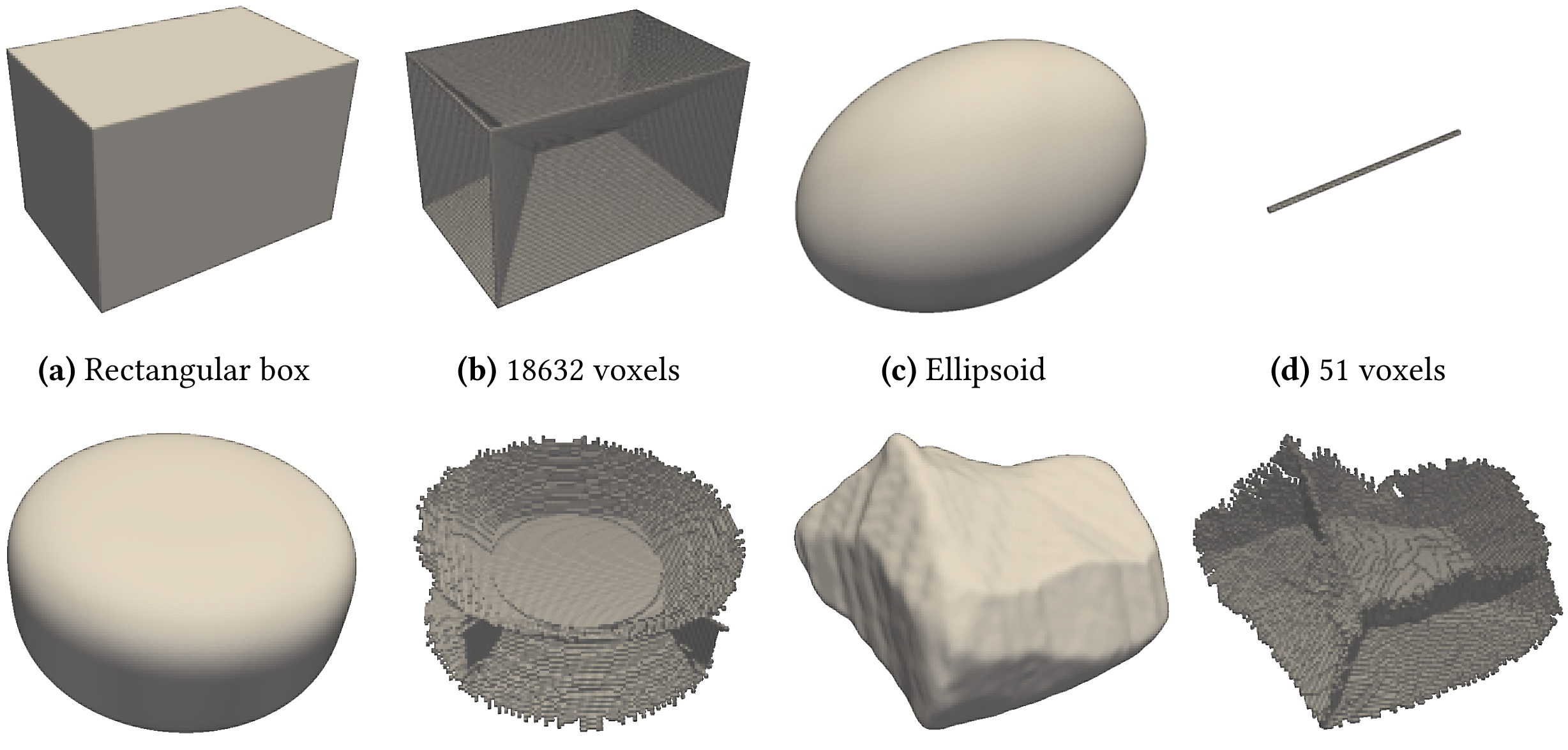

**(a)** Rectangular box **(b)** 18632 voxels **(c)** Ellipsoid **(d)** 51 voxels

**(e)** Super-ellipsoid **(f)** 16379 voxles **(g)** Irregular shape **(h)** 19824 voxels

**Figure 3:** Surface skeletons of four different shapes. Voxel size is set as $\frac{d_{eq}}{100}$ with $d_{eq}$ being the equivalent diameter of input mesh.

In most cases, surface skeletons contain a certain number of spurious skeletal parts (see Figure 3f and 3h), because of the nature of noise sensitivity in 3D thinning algorithms (even in surface-mesh based skeletonization schemes). Skeleton pruning methods are often applied to prevent such spurious skeletal parts appearing, however, not implemented in this work, because a surface skeleton is used to generate candidate spheres. Moreover, the number of spurious skeletal voxels is usually a small portion of the total voxels on the surface skeleton, and most generated candidate spheres on which are discarded by the greedy algorithm as described in the following section.

## 3. Realistic particle approximation

Now we are ready to use the medial surface (surface skeleton) of a given particle to approximate its shape with multiple overlapping spheres.

The first step is to generate all candidate spheres based on the medial surface made of voxels. For each skeletal voxel, a candidate sphere (i.e., maximal inscribed sphere, named *medial sphere*) is generated such that it is centred on the voxel and tangential to the particle' surface mesh on the inside. In practice, the radius vector of which is obtained by the skeletal voxel center to a point (vertex) on the particle's surface mesh, such that all other points are outside the candidate sphere. If the points on the surface mesh are dense enough, this sphere will be approximately tangential to the particle's inner surface.

Next, we want to select candidate spheres as few as possible to compose a clump, until it covers a certain percentage (e.g. 90%) of the volume of the original particle. Rather than using some user-defined parameters such as sphere-to-sphere distance and minimum radius that bring uncertainty, the greedy algorithm is utilized in the sphere-inserting process, in order that each sequentially inserted sphere has the greatest contribution to the *volume coverage*, which is defined by the ratio of the clump's volume to the particle's original volume. Thus with a desired clump volume coverage, i.e., the approximation accuracy for a particle, a minimum number of spheres per clump can be guaranteed.

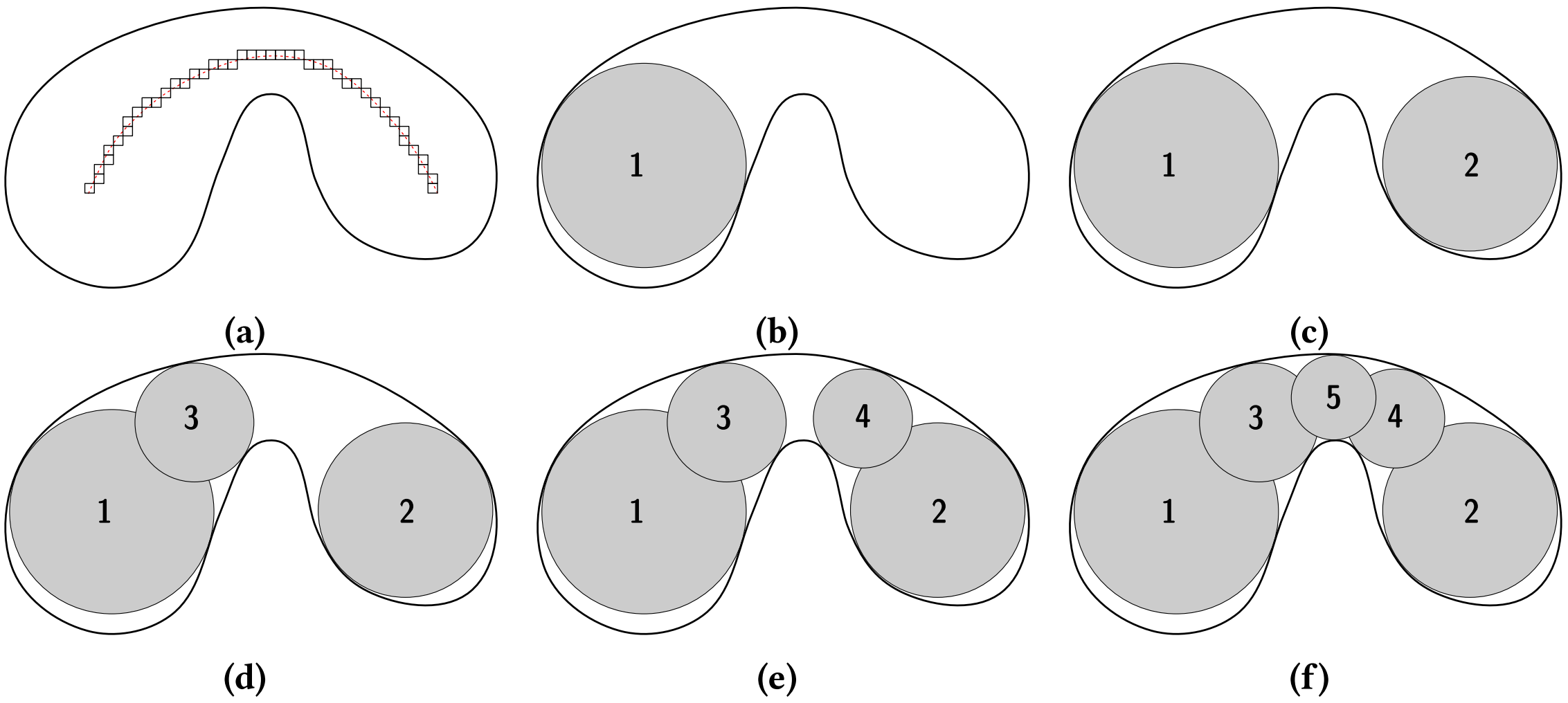

(a) (b) (c) (d) (e) (f)

**Figure 4:** Schematic operation of greedy algorithm: (a) particle's boundary and its skeleton made of voxels; (b-f) sequentially inserted spheres with greatest effective coverage.

A schematic of such greedy approach is illustrated in Figure 4. The criterion in the sphere-inserting process is based on the greatest *effective coverage*, which is defined by the sum of volumes of *uncovered* voxels within a candidate sphere. A voxel is said to be uncovered if it is not covered by any previously inserted spheres. Initially all candidate spheres are sorted into decreasing order of radius, thus the first inserted sphere (Figure 4b) is the one with greatest radius since all voxels are not covered by any spheres yet. Next, effective coverages of the rest candidate spheres need to be updated, so that the next sphere to be inserted can be found by selecting the one with greatest effective coverage. This operation is repeated until the termination condition is met.

Note that a clump made of spheres may not be continuous in the early stage of the sphere-inserting process as shown in Figure 4c, 4d and 4e. Nevertheless, up to a minimum volume coverage (e.g. $\geq 90\%$) to the original particle, spheres inside the clump are usually connected (see Figure 4f) in most cases depending on the particle's shape. Moreover, the clump's surface will become smoother and smoother as the volume coverage increases.

### 3.1 Discretization of particle's body with non-uniform grid

The most computationally intensive operation in the sphere-inserting process is updating the effective coverage of every candidate sphere at each iteration. It has a time complexity of $\mathcal{O}(n_1 \cdot n_2)$, where $n_1$ is the number of candidate spheres and $n_2$ is the number of uncovered cells (or voxels if all cells are uniform). $n_2$ is often several order of magnitude larger than $n_1$, if we use the same uniform grid in the 3D thinning and sphere-inserting processes. For example, the particle shown in Figure 3g is discretized into c.a. 0.52 million cells, while the computed surface skeleton has only 19824 cells. A coarser uniform grid can be used in order to speed up the process, but the evaluation of effective coverage for each candidate sphere, as well as the particle's shape approximation with coarser cells will be less accurate.

A possible solution to reduce the number of cells $n_2$ while keeping the approximation accuracy, is using finer border cells to capture a particle's shape, and coarser interior cells to fill the particle's body. In this way, much less cells are needed to mimic the particle's shape and volume.

Take the particle in Figure 3g as example. Let $d_{eq}$ be the equivalent diameter of the particle, and $N$ the parameter to decide the cell size: $d_{eq}/N$. Here $N_{out}$ and $N_{in}$ represent the parameters to decide the sizes of boundary and interior cells, respectively. As shown in Figure 5a, the particle's shape are in general well captured ($N = 100$) with blocky surface. However, because of the large number of cells (523624), it is quite time-consuming to update the effective coverage for every candidate sphere at each iteration. If fine cells ($N = 100$) are only used for boundary layers, and coarse cells ($N = 50$) for the interior, the number of cells can be dramatically reduced. Comparing the cross-sections of both grids in Figure 5b and 5c, it is clear that the non-uniform grid of 177577 cells has the same level of shape/volume approximation as the fine uniform grid.

Non-uniform grid such as the one shown in Figure 5c can be simply produced within the 3D surface thinning process. Let $B$ denote the set of border cells for $X$ with $N = 50$, and $B'$ the set of cells $\in \overline{X}$ that are adjacent to $B$, plus interior cells that are adjacent or within certain distance (e.g.

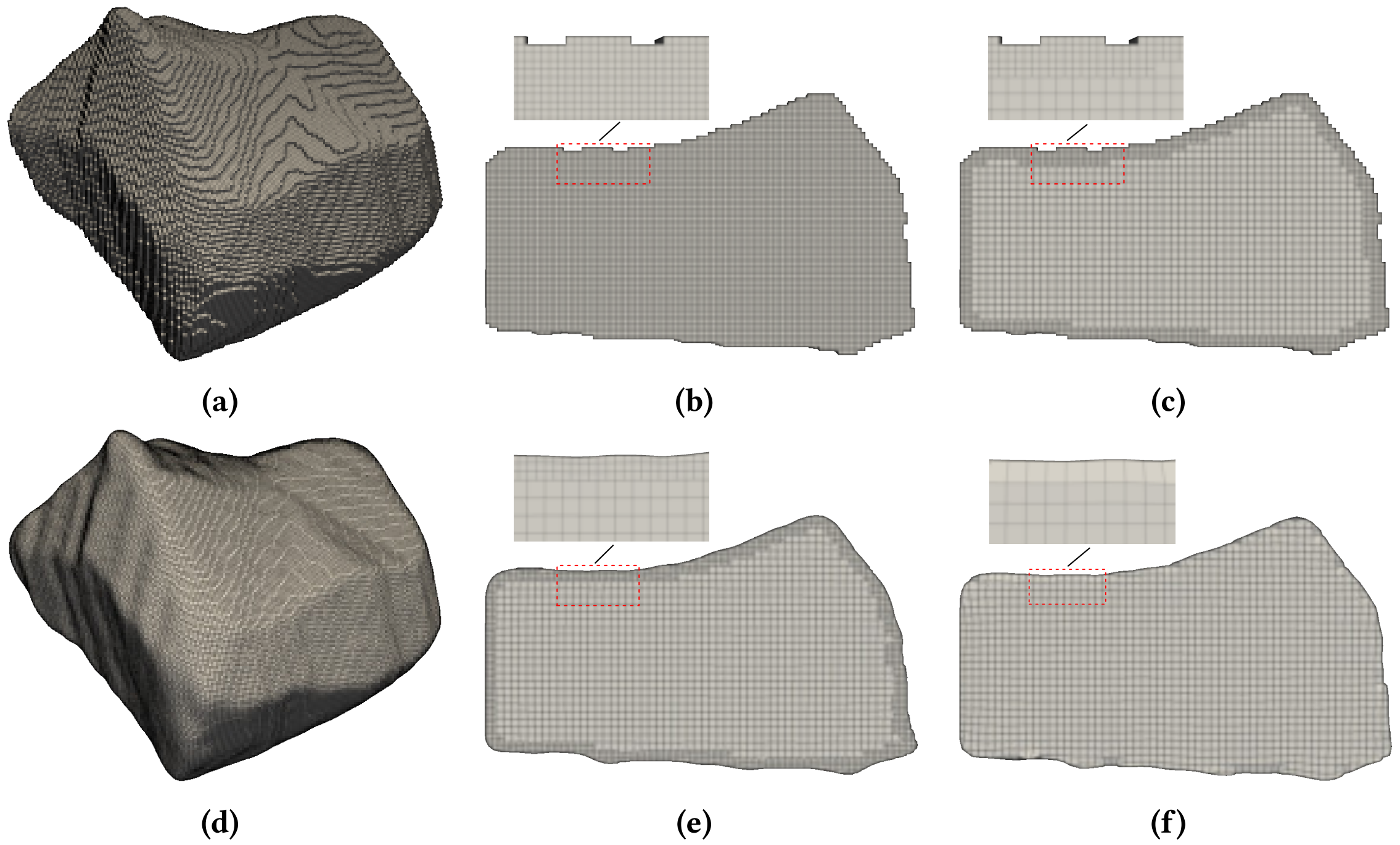

**Figure 5:** Particle discretization with uniform and non-uniform cells: (a) uniform grid, $N_{out} = N_{in} = 100$; (b) cross-section of (a); (c) cross-section of non-uniform grid with $N_{out} = 2N_{in} = 100$; (d) surface-conforming grid, $N_{out} = 2N_{in} = 100$; (e) cross-section of (d); (f) cross-section of surface-conforming grid with $N_{out} = N_{in} = 50$.

2-5 cell size) to the border cells. Here $B \in B'$, and $X$ is updated: $X = X \setminus B'$. If each cell in $B'$ is subdivided into 8 smaller unit cells, a new set $B'$ is formed with $N = 100$. After removing the cells in $B'$ that are outside the particle, a non-uniform grid is finally obtained by the set $B' \cup X$.

An even better way for the particle's discretization is to use a surface-conforming grid that could yield a smooth and accurate shape approximation, depending on the cell size on the boundary. Here an open-source meshing tool snappyHexMesh [23] is utilized to generate such grid. By re-meshing the non-uniform grid shown in Figure 5c with surface-conforming grid, the surface of the discretized particle (Figure 5d and 5e) is almost identical to the original particle mesh (Figure 3g). Without refinement of boundary cells, a coarser (Figure 5f) grid with 65485 cells, which is similar to the uniform grid, can be used to further speed up the clump generation process with slightly less accuracy.

## 3.2 Greedy set-covering

Despite the number of cells for a discretized particle can be significantly reduced by using fine cells only on the boundary (compare Figure 5c and 5e with Figure 5b), the computation of effective coverage is still expensive as the running time of each iteration is of $\mathcal{O}(n_1 \cdot n_2)$. Nevertheless, the whole clump generation process by sequentially inserting a sphere with maximum effective coverage, can be converted to a greedy set-covering problem (SCP) [37], which has a polynomial-time $(ln|X| + 1)$ complexity [10] with $X$ being a finite set that contains all cells of a discretized particle.

Given an arbitrarily-shaped particle, once its candidate spheres and secondary non-uniform grid (e.g. Figure 5c or 5e) are computed, an instance $(X, \mathcal{S})$ of the set-covering problem can be constructed. Here $X$ is a finite set and $\mathcal{S}$ is a family of subsets of $X$. If the non-uniform grid (representing particle's shape) consists of $n_X$ cells, then $X = \left\{c_1, c_2 \cdots c_{n_X}\right\}$ with each cell being an element. For each candidate sphere, a subset $s_i \subseteq X$ can be generated ($i = 1, 2, \cdots m$), such that it contains all cells inside this sphere. Thus $\mathcal{S} = \{s_1, s_2, \cdots s_m\}$ with $m$ being the number of candidate spheres. Assuming $X = \bigcup_{s_i \in \mathcal{S}} s_i$, now the problem is to find a minimum-size subset $\mathcal{C} \in \mathcal{S}$ such that $X = \bigcup_{s \in \mathcal{C}} s$, i.e., all cells in $X$ are covered by the members of $\mathcal{C}$.

A popular solution to solve set-covering problem is the greedy approach, i.e., picking a set with maximum number of uncovered elements (cells) at each iteration. However, this set does not necessarily have the maximum effective coverage among other sets, because of non-uniform grid used. Instead, the set $s$ with maximum weight (effective coverage) is selected and added in the minimum-size subset $\mathcal{C}$. Let $w(s)$ denote the weight of subset $s$, thus $w(s)$ can be calculated by $\sum_{i=0}^{k} v_i$, here $v_i$ is the volume of $i^{th}$ cell and $k$ is the number of cells in $s$. Next, the rest sets are updated by removing cells that are contained in $s$. The process is repeated until the set of maximum weight is zero at certain iteration. At this point, $X$ is fully covered by the minimum-size subset $\mathcal{C}$.

**Algorithm 2:** Greedy set-covering scheme

**Input:** Cell set $X$, set of candidate spheres $S$

**Output:** Minimum-size subset $\mathcal{C} \in \mathcal{S}$ that covers $X$

```
1  function greedySetCover(X, S)
2      𝒮 = generateSets(X, S);
3      U = X;
4      𝒞 = ∅;
5      while U ≠ ∅ do
6          select an s ∈ 𝒮 that has maximal weight;
7          U = U \ s;
8          𝒮 = 𝒮 \ {s};
9          updateSets(s, 𝒮);
10         𝒞 = 𝒞 ∪ {s};
11     return 𝒞;
```

Following the strategy proposed above, the pseudo-code for solving the greedy set-covering problem is described in Algorithm 2. The function generateSets$(X, S)$ is responsible for the generation of subsets of $X$ for all candidate spheres in $S$. After a subset $s$ with maximal weight being selected, the function updateSets$(s, \mathcal{S})$ removes cells $\in s$ from any subsets in $\mathcal{S}$.

As solving a greedy set-covering problem is performed in a polynomial-time $(ln|X| + 1)$, the actual clump generation process (line 5-10 in Algorithm 2) usually takes less than 30 seconds. The only expensive part is the sets generation (line 2), which takes up to c.a. 250 seconds on a single CPU core (2.5 GHz), depending on the number of candidate spheres $m$ and cells $n_X$. Since a particle's surface skeleton is computed in $\mathcal{O}(n)$ time, the whole particle shape approximation process (Algorithm 1 and 2) with a given particle's surface mesh as input and the clumps of coarse to fine resolutions as output, can be finished within few minutes. If we implement the algorithms with parallel computing API like OpenMP, the performance can be further improved.

### 3.3 Particle shape approximation

It is important to note that not all cells in $X$ are guaranteed to be covered by the candidate spheres, because the medial surface (surface skeleton) computed by the 3D thinning Algorithm 1 is an approximation of the analytical solution, given the fact that uniform cells are used to represent a particle's shape in the 3D thinning process. Therefore, those cells whose centers are not contained in any candidate spheres should be excluded from the cell set $X$ in Algorithm 2, we denote the new cell set as $X_{cs}$ (cells covered by all candidate spheres).

Here we approximate the shape of the flat particle (FP) in Figure 3g as an example. Let $V_p$ denote the volume sum of all cells in $X$. In this work, surface-confirming grid (e.g. Figure 5e) with $N_{out} = 2N_{in} = 100$ is used by default to represent a particle's volume and shape for solving the greedy set-covering problem. With this configuration, $V_p$ is c.a. 99.97% of the original volume of input mesh, which indicates that the cell set $X$ is accurate enough to represent the original particle shape.

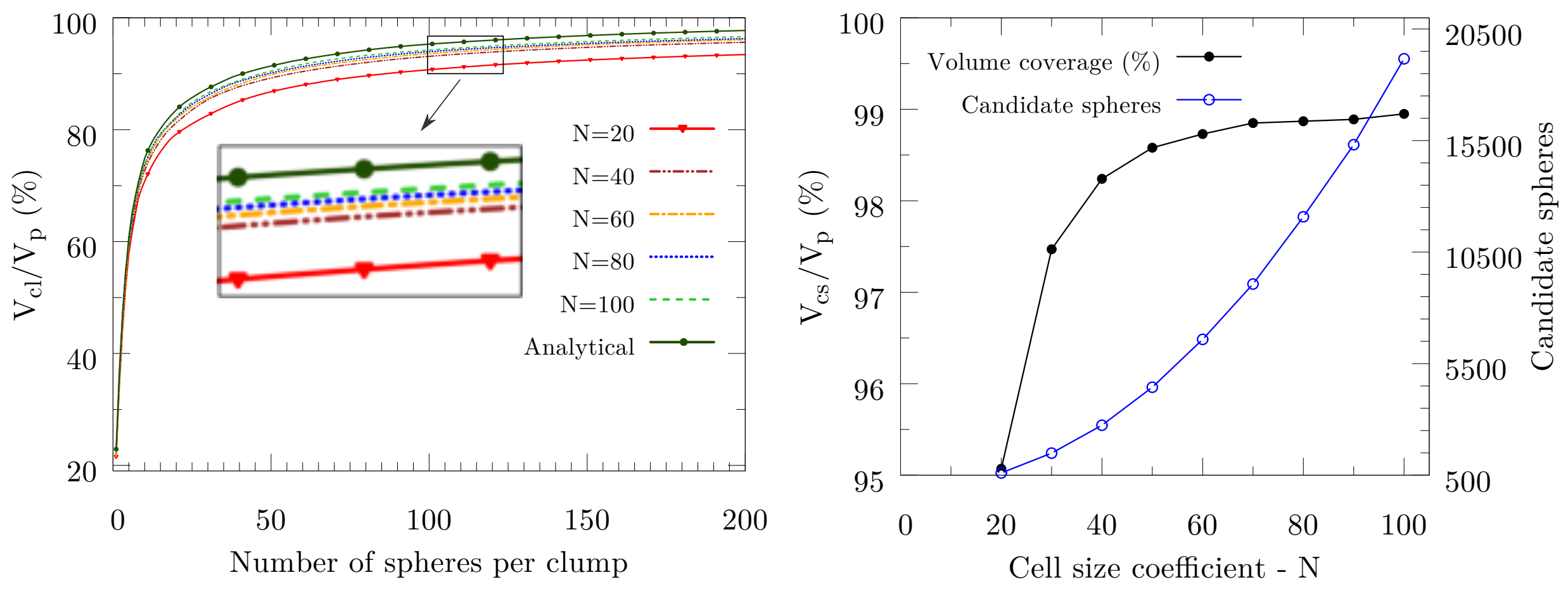

**Figure 6:** Volume coverage of the flat particle (FP): (left) clump volume coverage against spheres per clump with medial surfaces of different resolution; (right) number of candidate spheres and the volume coverage $V_{cs}/V_p$ with $N = 20 - 100$.

Let $V_{cs}$ denote the volume sum of all cells in $X_{cs}$. With increasing number of candidate spheres (equal to the number of skeletal voxels) computed in the 3D thinning process (voxel size is $d_{eq}/N$ with $N$ = 20 to 100), the volume coverage $V_{cs}/V_p$ increases rapidly from $N$ = 20 to 50. However, with $N$ ¿ 50, increased number of candidate spheres ($4.5 \times 10^3$ to $1.9 \times 10^4$) contributes less to the volume coverage as shown in Figure 6 (right). This implies that we do not have to compute the medial surface with a very fine cell resolution which leads to a large number of candidate spheres. For example at $N$ = 60, the number of candidate spheres is about $6.8 \times 10^3$ with a volume coverage of 98.8%; while at $N = 100$, a larger number of $1.9 \times 10^4$ candidate spheres covers nearly the same volume (99.0%) of $X$. Nevertheless, the center of a sphere with maximal weight is more likely closer to the analytical medial surface because of denser skeletal voxels, thus less spheres are required to compose a clump with a given clump volume coverage $V_{cl}/V_p$. Here $V_{cl}$ is the the volume sum of all cells covered by the generated clump.

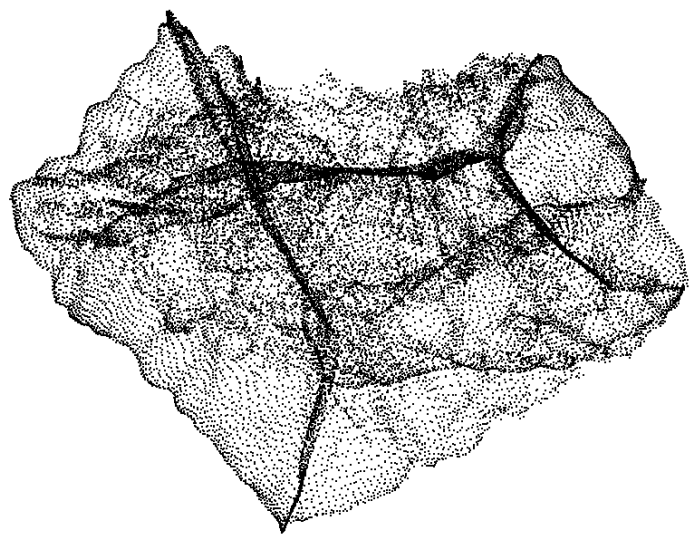

**(a)** 35305 skeletal points

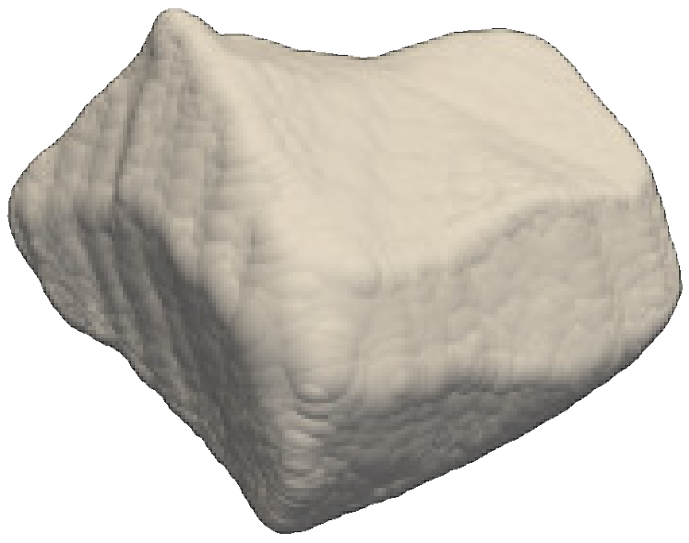

**(b)** 1565 spheres (100%)

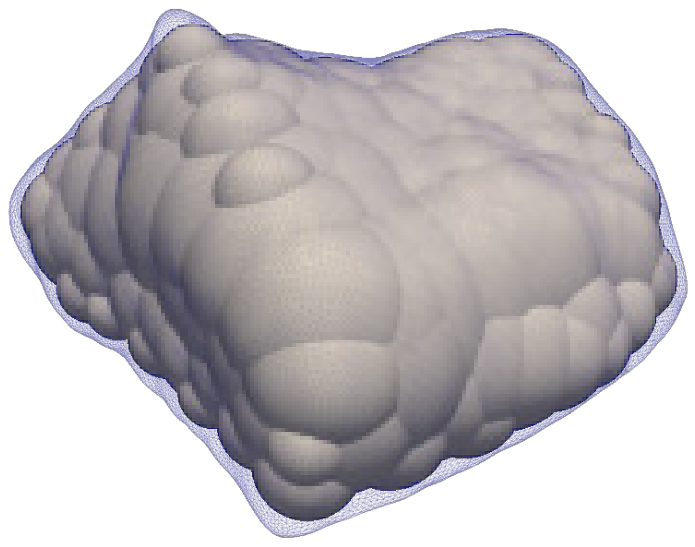

**(c)** 100 spheres (94.0%)

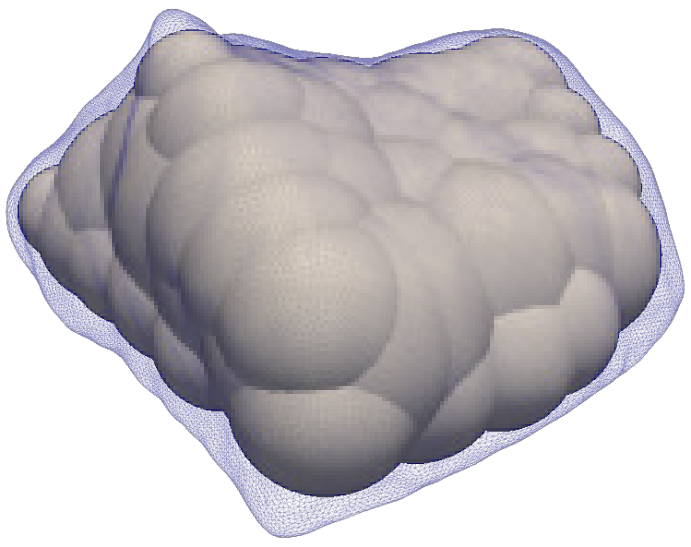

**(d)** 50 spheres (90.2%)

**Figure 7:** Multi-sphere approximation of the flat particle (Figure 3g) with decreasing volume coverage: (a) medial sphere centers calculated from surface points; (b) SCP result from the analytical medial surface shown in (a); (c-d) clump of 100, 50 spheres at $N = 100$.

Comparison of the clump approximation accuracy, i.e., the clump volume coverage $V_{cl}/V_p$ against the number of spheres per clump is plotted in Figure 6 (left) between different settings of medial surfaces. The results show that the clump computed with a medial surface of higher resolution has larger volume coverage for a fixed number of spheres per clump. The “analytical” medial surface (made of points) as shown in Figure 7a is simply calculated from the surface points (35305) of the FP input mesh: for each surface point, a skeletal point (center of maximal inscribed sphere) on the internal point normal can be obtained, on which a medial sphere centred only contains the forming surface point and another point on the mesh [18]. If the surface points on the mesh is dense enough, the computed medial surface is considered to be a close approximation of the analytical one. Out of the 35305 candidate spheres, only 1565 spheres (Figure 7b) are chosen to cover all cells in $X$ by the greedy set-covering algorithm; for $N = 60 - 100$, the SCP result $C$, which covers around 99% volume of $X$, contains c.a. 2000 spheres.

For different clump volume coverage such as $90\%, 85\%$ and $80\%$, the number of spheres per clump are 41, 24 and 15 in the analytical solution; while for medial surfaces computed by the 3D thinning algorithm at $N = 100, 80, 60$, the numbers are slightly larger: 49, 50 and 51 at 90% coverage; 27 at 85% coverage and 17 at 80% coverage for all the three cell size coefficients. It is clear that the difference in number of spheres per clump between different cell/voxel size coefficients

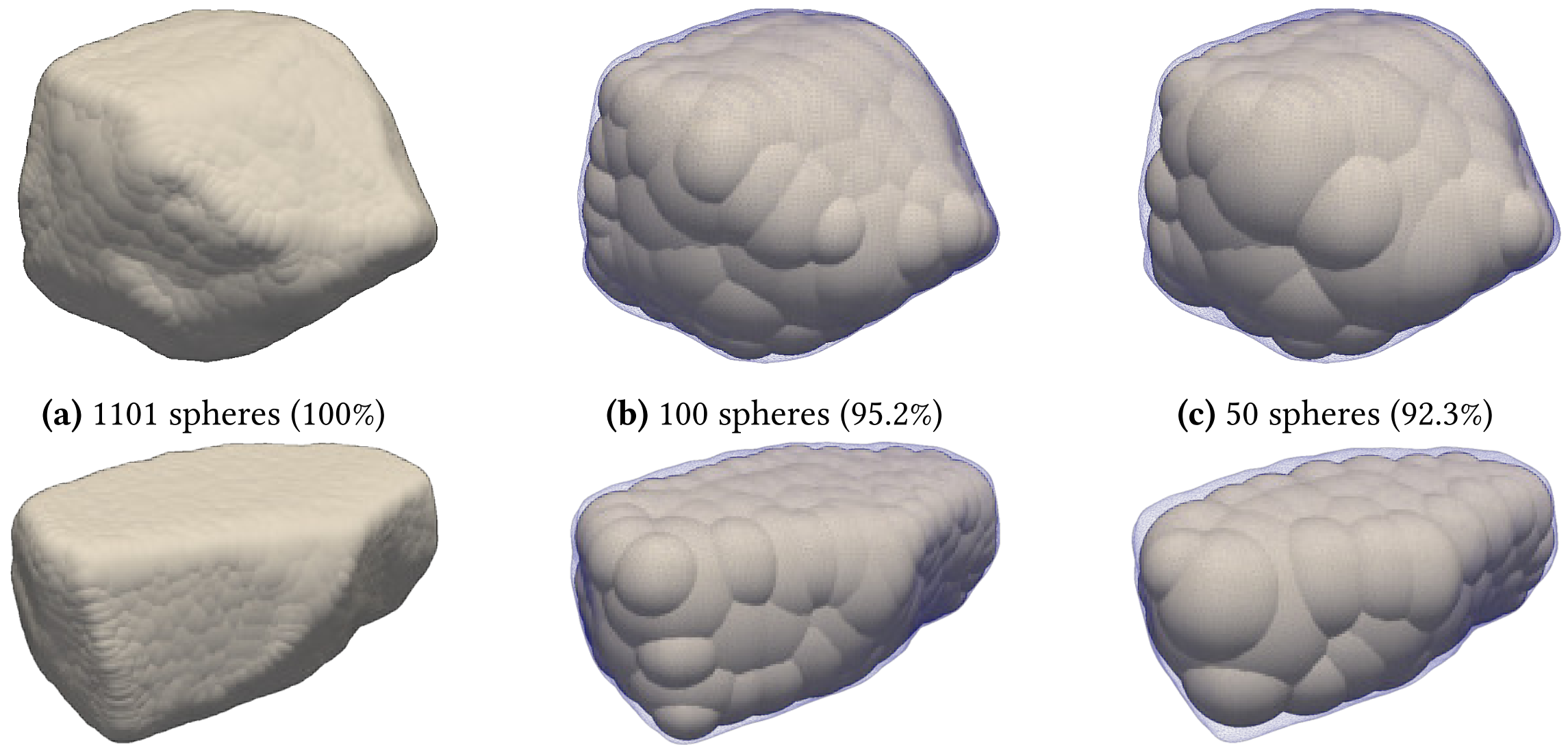

**(a)** 1101 spheres (100%) **(b)** 100 spheres (95.2%) **(c)** 50 spheres (92.3%)

**(d)** 1376 spheres (100%) **(e)** 100 spheres (95.0%) **(f)** 50 spheres (91.1%)

**Figure 8:** Multi-sphere approximation of compact particle (CP) and elongated particle (EP): (a) analytical SCP result of CP; (b-c) clump of CP with 100, 50 spheres at $N = 100$; (d) analytical SCP result of EP; (e-f) clump of EP with 100, 50 spheres at $N = 100$.

is marginal, therefore $N = 60$ is suggested for use in the 3D thinning process, as it takes only 16 seconds to compute the surface skeleton (6806 voxels), and the number of candidate spheres can be significantly reduced, as a result the whole clump generation can be finished in less than a minute. Furthermore, if a particle were approximated at coarse level (e.g. 85% and 80% volume coverage), the needed amount of spheres per clump is very close to the analytical solution. It implies that a coarse medial surface ($N = 60$) is adequate for the coarse clump approximation.

Note that if candidate spheres (set $S$) are generated on the cells centers of the coarse surface-conforming mesh shown in Figure 5f, $X$ (i.e., Figure 5e) can be fully covered. However, the clump approximation accuracy ($V_{cl}/V_p$) is close to the case of $N = 50$ with significantly less candidate spheres (4556), whereas the whole clump generation process takes much longer (c.a 10 minutes) than its counterpart (40 seconds). It implies that extra candidate spheres whose centers are not located on the medial surface do not contribute to the approximation accuracy, but make the sphere-inserting process much more expensive in terms of computation and memory usage.

The particle shape has a significant impact on the number of spheres per clump (denoted by $n_{cl}$) for a given approximation accuracy (i.e., clump volume coverage). Figure 8 shows another two shapes and their multi-sphere approximation. The *sphericity* $\Psi$ of a particle, calculated by $\pi^{\frac{1}{3}}(6V_p)^{\frac{2}{3}}A_p^{-1}$ where $A_p$ is the surface area of the particle, is used to describe the particle's shape factor here. Let us compare the sphericity and its influence on the multi-sphere approximation for three different types of particle shape: namely the compact particle (CP) in Figure 8a, the elongated particle (EP) in Figure 8d, and the flat particle (FP) in Figure 7b. With decreasing sphericity ($\Psi_{CP} = 0.91$, $\Psi_{EP} = 0.85$, $\Psi_{FP} = 0.84$), we can clearly see that the number of spheres, i.e., the solution of SCP for analytical medial surface, to fully cover all cells in $X$ increases. That means for particles with higher sphericity (compactness), less spheres are required to compose a clump for a given level of approximation accuracy.

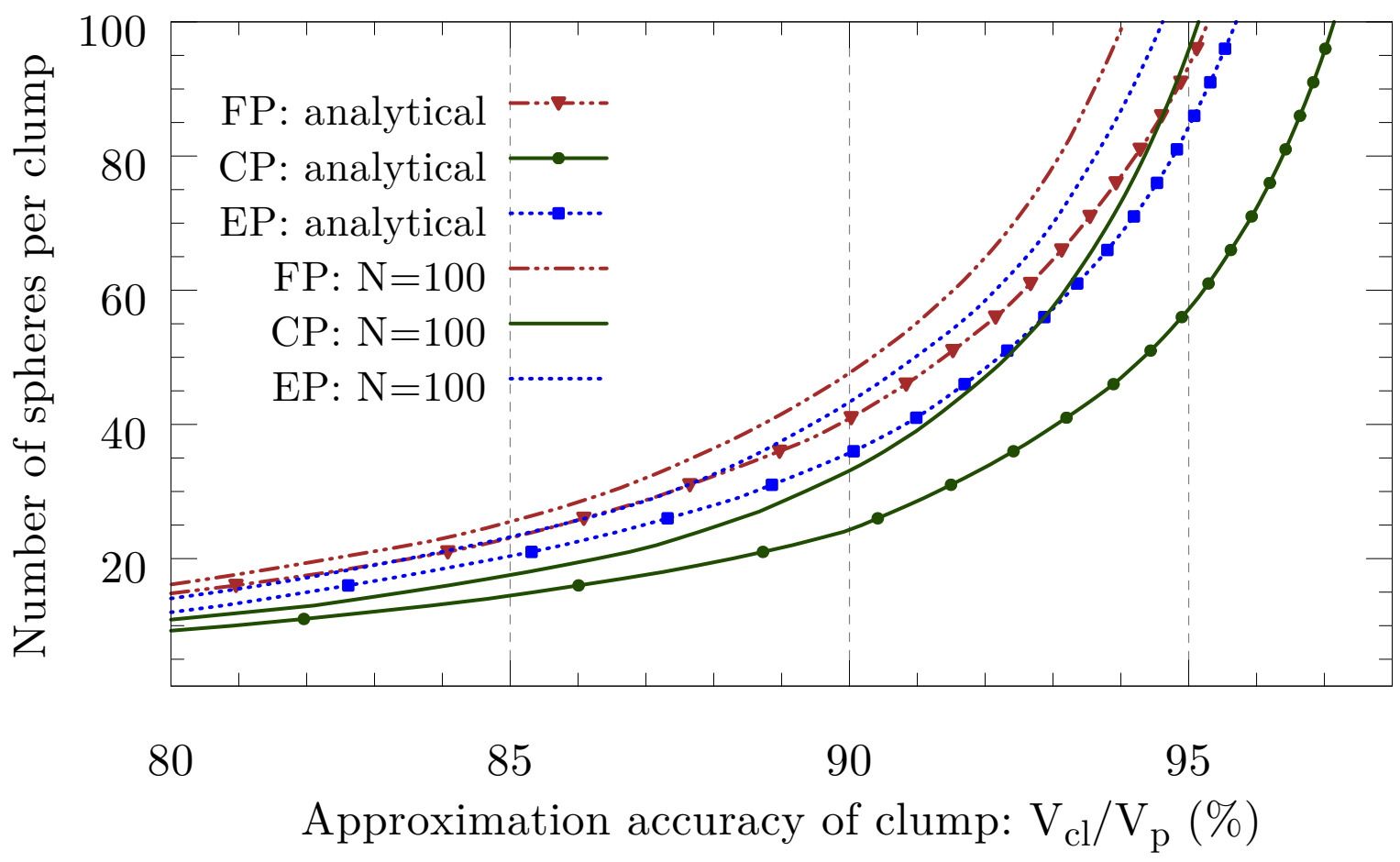

**Figure 9:** Volume coverage vs. spheres per clump of three different particle shapes.

To model a granular system with large number of particles, we want the generated clumps contain as few spheres as possible in order to speed up the simulation, while keeping a minimum approximation accuracy. Here we are particularly interested in the clumps with less than 100 component spheres. Assuming the coarse level of clump approximation is 90%, and the intermediate level 95%; let $n_{cl}^{90}$ and $n_{cl}^{95}$ denote the number of spheres per clump for the coarse and intermediate level, respectively. As shown in Figure 9, the analytical solutions of $n_{cl}^{95}$, which are inversely proportional to the sphericity of the prototypical particle (i.e., the input mesh), are 58, 85, 93 spheres for the three types of particle shape CP, EP and FP, respectively; both of them are under the 100-sphere mark. While at $n_{cl} = 100$, the clump volume coverages of EP and FP at $N = 100$ are 94% and 94.6%, which are still close to the intermediate level mark - 95%. If particles were modelled at coarse level, a significantly less number of spheres per clump ($n_{cl}^{90}$) is required: 25, 36, 41 spheres (CP, EP and FP respectively) to achieve 90% volume coverage in the analytical solution, and slightly larger numbers 33, 43, 48 with medial surfaces being computed by the 3D thinning algorithm at $N = 100$.

## 4. Mechanical properties of generated clumps

Assuming the mass distribution of real particles is continuous with uniform density $\rho$, the mass $M$, center of mass $\boldsymbol{C}$ and inertia tensor $\boldsymbol{I}^C$ of a generated clump can be calculated via all cells covered by the clump: $M = \sum_{i=0}^{n} m_v$, $\boldsymbol{C} = \frac{1}{M}\sum_{i=0}^{n} m_v \boldsymbol{r}_i$, and the momentum of inertia around the centre of mass

$$\boldsymbol{I}^C = m_v \begin{bmatrix} \sum_{i=1}^{n_v}(y_i^2 + z_i^2) & -\sum_{i=1}^{n_v}(x_i y_i) & -\sum_{i=1}^{n_v}(x_i z_i) \\ & \sum_{i=1}^{n_v}(x_i^2 + z_i^2) & -\sum_{i=1}^{n_v}(y_i z_i) \\ symmetric & & \sum_{i=1}^{n_v}(x_i^2 + y_i^2) \end{bmatrix} \quad (1)$$

where $n_v$ is the number of cells (or voxels), $m_v$ is the mass of unit cell, $\boldsymbol{r}_i$ and $(x_i, y_i, z_i)$ are the global and local (relative to $\boldsymbol{C}$) position vector of $i^{th}$ cell center. These properties are simply obtained by discretizing the clump into very fine uniform cells as shown in Figure 1, and each cell is considered as a point mass instead of a finite-size cube for simplicity. Here a uniform grid with $N = 200$ (3.5 - 4.2 million unit cells depending on the number of spheres per clump) is used for an accurate approximation.

The pre-calculated aggregate properties $M$, $\boldsymbol{C}$ and $\boldsymbol{I}^C$ can be explicitly assigned to each clump, in order to avoid the error in computing these three attributes via highly overlapped spheres, if each component sphere has the same density or same mass. Nevertheless, this may be quite tedious or computationally expensive (aggregate properties calculated in DEM program) for a system with large number of particles of different shapes.

A better alternative is to project these properties onto the component spheres of a clump by setting each with different density, such that the aggregate properties of the component spheres are identical to the prototypical clump. In this way, each particle template (prototypical clump), in which each component sphere is given by its center, radius and density, can be read into program without explicitly setting or calculating the inertia, mass and center of mass.

If a clump is composed of $n_{cl}$ solid spheres, a linear system of 10 equations can be formed as follows in order to correct the mechanical properties.

Momentum of inertia:

$$\begin{aligned} \textstyle\sum_{i=1}^{n_{cl}}(y_i^2 + z_i^2 + \tfrac{2}{5}r_i^2) \cdot m_i = I_{xx} \\ \textstyle\sum_{i=1}^{n_{cl}}(x_i^2 + z_i^2 + \tfrac{2}{5}r_i^2) \cdot m_i = I_{yy} \\ \textstyle\sum_{i=1}^{n_{cl}}(x_i^2 + y_i^2 + \tfrac{2}{5}r_i^2) \cdot m_i = I_{zz} \end{aligned} \tag{2}$$

$$\begin{aligned} -\textstyle\sum_{i=1}^{n_{cl}}(x_i y_i) \cdot m_i = I_{xy} \\ -\textstyle\sum_{i=1}^{n_{cl}}(x_i z_i) \cdot m_i = I_{xz} \\ -\textstyle\sum_{i=1}^{n_{cl}}(y_i z_i) \cdot m_i = I_{yz} \end{aligned} \tag{3}$$

in which the right-hand side are the diagonal and off-diagonal components of the inertia tensor $\boldsymbol{I}^C$. $r_i$ and $(x_i, y_i, z_i)$ are the sphere radius and the center relative to $\boldsymbol{C}$.

Center of mass:

$$\begin{aligned} \textstyle\sum_{i=1}^{n_{cl}} \frac{x_i'}{M} \cdot m_i = C_x \\ \textstyle\sum_{i=1}^{n_{cl}} \frac{y_i'}{M} \cdot m_i = C_y \\ \textstyle\sum_{i=1}^{n_{cl}} \frac{z_i'}{M} \cdot m_i = C_z \end{aligned} \tag{4}$$

where the right-hand side are the x-, y- and z-component of $\boldsymbol{C}$, $(x_i', y_i', z_i')$ is the global coordinate of the sphere center.

Total mass:

$$\sum_{i=1}^{n_{cl}} m_i = M \tag{5}$$

The following matrix equation is equivalent to the above equations in (2), (3), (4) and (5)

$$\boldsymbol{A}\boldsymbol{x} = \boldsymbol{b} \tag{6}$$

here $\boldsymbol{A}$ is a $10 \times n_{cl}$ matrix that stores coefficients of $m_i$, $\boldsymbol{x}$ is a column vector of the $n_{cl}$ unknowns $[m_1, m_2, \ \dots \ m_i, \ \dots \ m_{n_{cl}}]^T$. $\boldsymbol{b}$ is a column vector that stores the mechanical properties of the prototypical clump: $[I_{xx}, I_{yy}, I_{zz}, I_{xy}, I_{xz}, I_{yz}, C_x, C_y, C_z, M]^T$.

Obviously, there is no solution for $\boldsymbol{Ax} = \boldsymbol{b}$ if $n_{cl} < 10$. When $n_{cl} = 10$, we may find an exact solution, but the mass is not guaranteed to be positive. In most cases, the number of spheres $n_{cl}$ per clump is required to be greater than 10 for a better shape approximation. Therefore, the system of linear equations $\boldsymbol{Ax} = \boldsymbol{b}$ becomes indeterminate, i.e., there are infinite number of solutions. In order to obtain a solution with $m_i > 0$, the indeterminate system $\boldsymbol{Ax} = \boldsymbol{b}$ can be converted to a linear programming (LP) problem [10] as follows.

$$\text{maximize } \boldsymbol{fx}, \text{ subject to } \begin{cases} \boldsymbol{Ax} = \boldsymbol{b} \\ \boldsymbol{lb} \leq \boldsymbol{x} \leq \boldsymbol{ub} \end{cases} \tag{7}$$

here $\boldsymbol{f}$ is a row vector of coefficients, $\boldsymbol{fx}$ is the objective function, $\boldsymbol{lb}$ and $\boldsymbol{ub}$ represent the column vectors (same dimension as $\boldsymbol{x}$) containing the lower and upper bound on each of the unknowns $m_i$. In practice, $\boldsymbol{lb}$ and $\boldsymbol{ub}$ can be simply set as $M_{min}[1, 1, \dots 1]^T$ and $M_{max}[1, 1, \dots 1]^T$ respectively, where $M_{min} = \frac{M}{1000}$ and $M_{max} = M$. In order to form a simple object function, we can set $\boldsymbol{f}$ as, e.g. $[0, 0, \dots 0, 1]$ such that $\boldsymbol{fx} = m_{n_{cl}}$.

It is found that the least number of spheres per clump required for an exact solution varies from 23 to 48 for the 3 different particle shapes in this work. This is already good in terms of spheres per clump (¡= 50) for simulating a granular system of small to intermediate size. Nevertheless, for large systems with over 1 million particles, we want to further reduce the least number of spheres per clump for an exact solution. Since the center of mass and principal axes (eigenvectors of $\boldsymbol{I}^C$) of a clump are pre-calculated, we can first move the clump such that $\boldsymbol{C} = (0, 0, 0)$, then align the principal axes to the global X-, Y- and Z-axis. The translational and rotational transformation can be described mathematically: $\boldsymbol{X}^{new} = \boldsymbol{R}(\boldsymbol{X} - \boldsymbol{C})$, where $\boldsymbol{X}$ and $\boldsymbol{X}^{new}$ are the global position vector of a sphere center before and after the transformation, and $\boldsymbol{R}$ is the $3 \times 3$ rotation matrix for aligning the clump's principal axes to the global X-, Y- and Z-axis. At this point, we can use $\boldsymbol{X}^{new}$ to replace the sphere centres in equations (2), (3) and (4). Because the diagonal components of the inertia tensor are the principal moments of inertia, and off-diagonal components become zero after the rotation, thus the equations in (3) are not necessary. Therefore, a new linear system with 7 equations can be formed from (2), (4) and (5). Indeed, an exact solution for the new linear system also satisfies (3) even though not included.

In general, an optimal solution maximizing the objective function can be found for the 7-equation linear system, if the number of spheres per clump ($n_{cl}$) is between 15 and 30 depending on the particle shape. In other words, we can find an exact solution of the mass set $(m_1, m_2, \dots m_{n_{cl}})$ that satisfies the conditions (2), (4) and (5), hence each component sphere can be assigned with different density according to the radius. At this point, we can say that the mass distribution inside the clump is "corrected", because the aggregate properties directly computed via overlapped spheres are identical to the prototypical clump.

In some cases there is no solution for the LP problem (7) if the target clump is composed of few spheres. For example, if particles were approximated at 90% volume coverage, we are not able to find a mass set for the compact particle ($n_{cl}^{90} = 25$), such that the aggregate properties of these 25 spheres match the prototypical clump. However, if a small error $\varepsilon$ (e.g. $\leq 5\%$) is allowed for the inertia, we can reconstruct (7) as follows.

$$\text{maximize } \boldsymbol{fx}, \text{ subject to } \begin{cases} (1-\varepsilon)\boldsymbol{b}_1 \leq \boldsymbol{A}_1\boldsymbol{x} \leq (1+\varepsilon)\boldsymbol{b}_1 \\ \boldsymbol{A}_2\boldsymbol{x} = \boldsymbol{b}_2 \\ \boldsymbol{lb} \leq \boldsymbol{x} \leq \boldsymbol{ub} \end{cases} \tag{8}$$

where $\boldsymbol{A}_1$ and $\boldsymbol{b}_1$ are the matrices that store the coefficients of the left-hand side and right-hand side of the linear equations in (2), receptively; $\boldsymbol{A}_2$ and $\boldsymbol{b}_2$ are the matrices that store the coefficients of the left-hand side and right-hand side of the linear equations in (4) and (5), receptively. Note that $\boldsymbol{A} = [\boldsymbol{A}_1; \boldsymbol{A}_2]$, and $\boldsymbol{b} = [\boldsymbol{b}_1; \boldsymbol{b}_2]$. We can increase the error $\varepsilon$ from zero gradually until a solution is found for the LP problem (8). In the case of the compact particle approximated at 90% volume coverage, a solution is found when $\varepsilon = 0.7\%$.

Let $\varepsilon_w$ denote the weighted mean error of the principal moments of inertia: $\varepsilon_w = \frac{\sum_{i=1}^{3}(\lambda_i - \lambda_i')}{\sum_{i=1}^{3}\lambda_i}$, where $\lambda_i$ and $\lambda_i'$ are the $i^{th}$ component of the principal moments of inertia of the real particle and the prototypical clump (unit cells covered by component spheres), respectively. It is found that $\varepsilon_w$ is only related to the clump volume coverage: $\varepsilon_w$ is approximately equal to 16% at the coarse level, and 8.5% at the intermediate level, regardless of the particle shapes and the number of spheres per clump. In this sense, a small error ($< 5\%$) in the inertia has less significant impact on the dynamic behaviour of generated clumps, because the error in the shape approximation and inertia between the real particle and the prototypical clump (millions of unit cells covered by the component spheres) plays a major role on the dynamic behaviour. Nevertheless, we can slightly increase the number of spheres per clump such that an exact solution of (8) can be found.

Note that the LP problem (8) is identical to (7) when $\varepsilon = 0\%$. Therefore, (8) can be used to obtain either exact or approximated solutions for clumps with low number of spheres. As the object function is to maximize the mass of one of the clump's component spheres, most of the rest spheres tend to have masses close to the mass lower bound. In order to obtain a more evenly distributed mass set, we can add an extra variable $m_e$ to the end of $\boldsymbol{x}$, and $n_{cl}$ constrains to (8): $m_i > m_e$ where $i = 1, 2, \ldots n_{cl}$. By setting $\boldsymbol{fx} = m_e$ as objective function, the mass of every component sphere is considered maximized.

The LP problem (8) is solved using the revised simplex method offered by the GLPK (GNU Linear Programming Kit) package in the open-source mathematical toolbox GNU Octave [16]. As the GLPK is also shipped with an ANSI C callable library, we can integrate the LP solver into the previous algorithms. Normally it takes only a moment to find the optimum solution of (8), if the unknowns are less than 100, i.e., the generated clump is composed of less than 100 spheres. Thus the computational cost for correcting the clump's mechanical properties is negligible compared to the 3D thinning and greedy set-covering algorithms.

# 5. Test cases

Three test cases were carried out to showcase the multi-sphere models generated by the combined algorithm in real applications, and to further validate the DEM code Rigid3D with different type of particle shapes: M&M candy (rounded surface), bi-convex tablet (rounded surface + sharp edge), and Stanford bunny (irregular surface). The simulation setup for the test cases: particle-wall contact, M&M candy packing in a cylindrical container and the optimal clump resolution are identical to those in the SR-DEM paper, the difference here is that the particles are approximated by multi-sphere models in this work. For the sake of completeness so that this article can be read independently, the details about the simulation setup and material properties are repeated here.

## 5.1 Particle-wall contact

This test case is inspired by the work of Kodam et al. [34], and aims to verify the effect of the resolution of "sharp" edges on the particle-wall contact. The tablet geometry and the simulation setup are shown in Figure 10. A bi-convex tablet, oriented at a specified angle, impacts a flat wall with a prescribed translational velocity normal to the wall and zero angular speed.

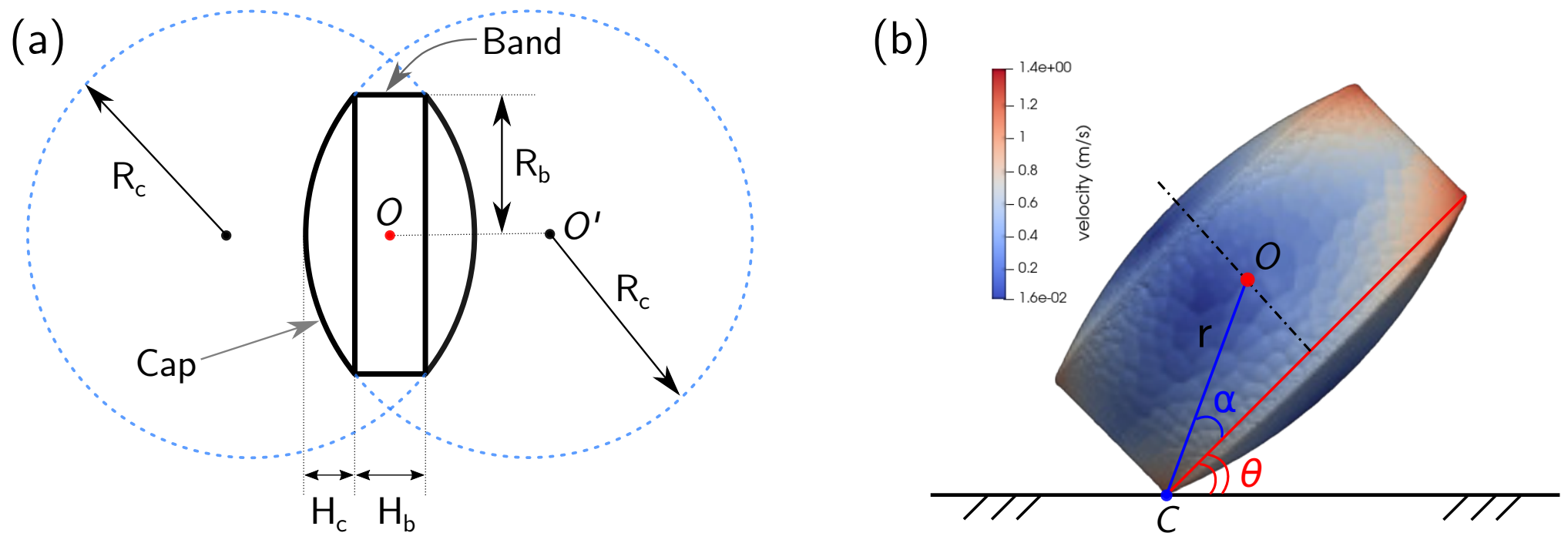

**Figure 10:** Particle-wall contact. (a) 2D schematic of a bi-convex tablet formed by two spherical caps and a cylinder; (b) schematic of particle-wall contact where the tablet is approximated by 2466 spheres.

In the simulation the particle-wall contact is assumed to be frictionless, and tablet's gravity is neglected. The post-impact angular and translational velocities can be written as

$$\omega_y^+ = \frac{mV_z^-(1+e)r_x}{I_y + mr_x^2} \tag{9}$$

$$V_z^+ = \omega_y^+ r_x - eV_z^- \tag{10}$$

where $m$ is the tablet mass, $e$ is the coefficient of restitution at the point of contact, $V_z^-$ is the pre-impact translational velocity (1 $m/s$), $\theta$ is the impact angle, and $I_y$ is the the y-component of principal moments of inertia. $r_x$ is the projection length of line segment $OC$ (length $r$) on X-axis, where $O$ is tablet center and $C$ is the contact point shown in Figure 10(b). Note that if the impact angle $\theta$ is less than a threshold, say $\theta^* = \sin^{-1}(R_b/R_c) \approx 24.5^\circ$, the impact occurs with the spherical cap, otherwise with the tablet's edge. Therefore $r_x$ reads

$$r_x = \begin{cases} h\cos(\theta) & \text{for } \theta < \theta^* \\ r\sin(\alpha + \theta) & \text{for } \theta \geq \theta^* \end{cases} \tag{11}$$

here $h$ is the distance from the tablet center $O$ to the center of the sphere $O'$ forming the cap shown in Figure 10(a). $h = R_c - (H_c + H_b/2)$, $r$=$\sqrt{(R_b^2 + H_b^2/4)}$ and $\alpha$ =$\tan^{-1}(\frac{H_b}{2R_b})$. Tablet's material properties and simulation parameters are listed in Table 1.

**Table 1:** Parameters used in particle-wall contact simulation

| Parameter | Value |
|---|---|
| Tablet radius $R_b$ (m) | $5.675 \times 10^{-3}$ |
| Tablet band height $H_b$ (m) | $4.0 \times 10^{-3}$ |
| Tablet cap height $H_c$ (m) | $1.23 \times 10^{-3}$ |
| Volume (m$^3$) | $5.312 \times 10^{-7}$ |
| Mass (kg) | $6.328 \times 10^{-4}$ |
| Moments of inertia $I_x$, $I_y$ (kg m$^2$) | $6.231 \times 10^{-9}$ |
| Moments of inertia $I_z$ (kg m$^2$) | $9.396 \times 10^{-9}$ |
| Shear modulus (GPa) | 1.15 |
| Poisson's ratio | 0.3 |
| Coefficient of friction | 0.0 |
| Coefficient of restitution | 0.6 |
| Time step $\Delta t$ (s) | $1.0 \times 10^{-7}$ |

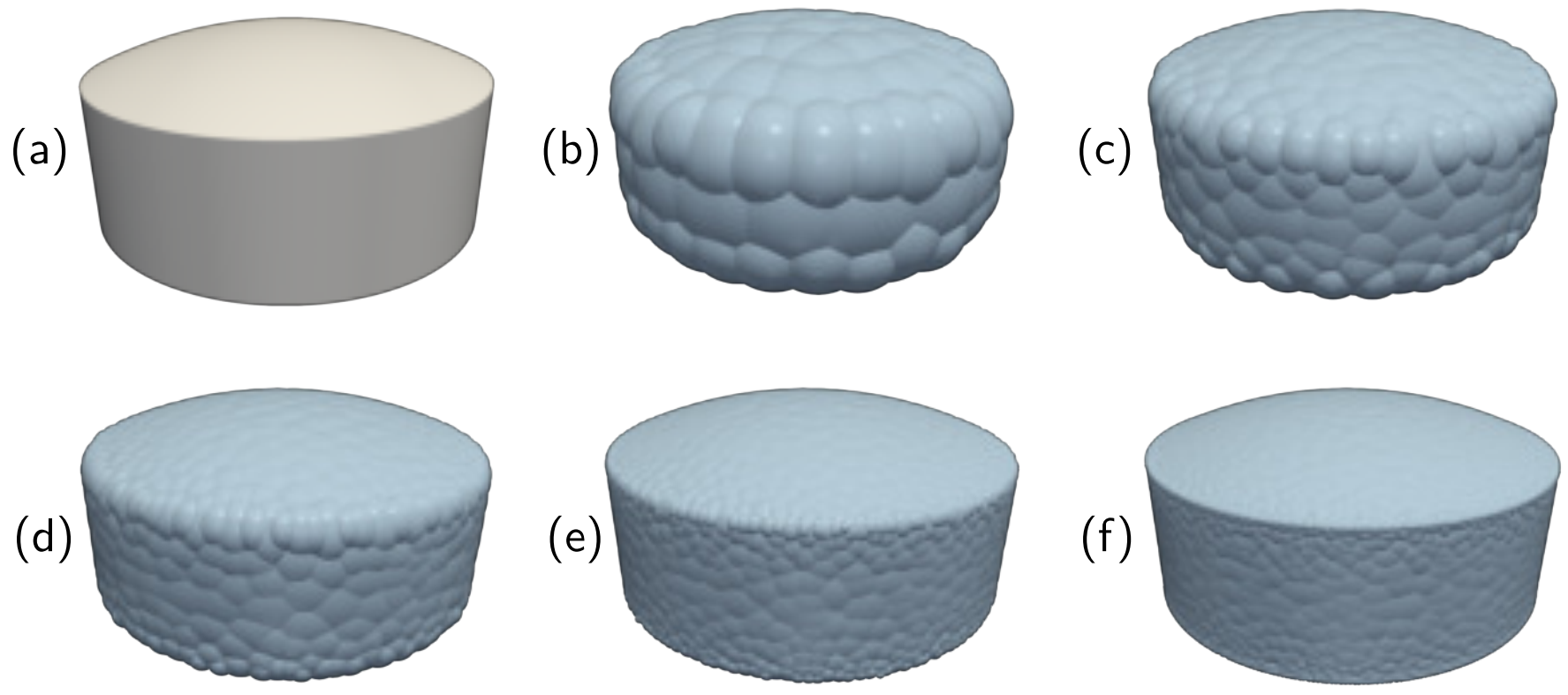

**Figure 11:** Bi-convex tablet MS models with increasing resolution in number of spheres and volume coverage. (a) original geometry; (b) 96 spheres, 95.04%; (c) 250 spheres, 97.85%; (d) 500 spheres, 99.13%; (e) 1000 spheres, 99.77%; (f) 2466 spheres, nearly 100% coverage.

The multi-sphere models for the bi-convex tablet are plotted in Figure 11 with increasing resolution, and are set to have the same mass and momentum of inertia as the true tablet for simplicity. Alternatively, we can discretize the multi-sphere models into millions of small uniform cubes, such that the sum of all cubes mass is the same as the original tablet. Then, the center of mass, and momentum of inertia can be computed upon all the cubes. Normally coarse models have larger deviation on these values than those with higher resolution (i.e. better shape approximation).

We select three levels of clump's volume coverage: 95%, 99%, and 99.9% to represent the coarse, intermediate and fine resolution, with the corresponding component spheres 96, 456 and 1329, respectively. Nevertheless, the finest resolution, 100% volume coverage with 2466 component spheres, are chosen to examine the accuracy of the "sharp" edge in the simulation. We notice that the resolution from 99.9% to 100% volume coverage, i.e. only 0.1% increase in the approximation accuracy, the corresponding component spheres are almost doubled. It implies that a large number of spheres are only used to mimic the sharp edges. However, their contribution to the overall approximation accuracy is minimal.

The particle-wall contact is simulated at various orientation angles with an interval of 5°, i.e. $\theta = \{0^\circ, 5^\circ, 10^\circ, \cdots, 90^\circ\}$ with an extra angle at 87.5°. Figure 10(b) shows an instantaneous snapshot of the tablet surface velocity profile just a moment after the contact takes place. The velocity of any component sphere can be calculated by $\boldsymbol{V}_i = \boldsymbol{V}_{cm} + \boldsymbol{\omega} \times \boldsymbol{r}_i$, where $\boldsymbol{r}_i$ is the position vector from the tablet's center to the component sphere's center, $\boldsymbol{V}_{cm}$ and $\boldsymbol{\omega}$ are the translational and rotational speed at the tablet's center of mass.

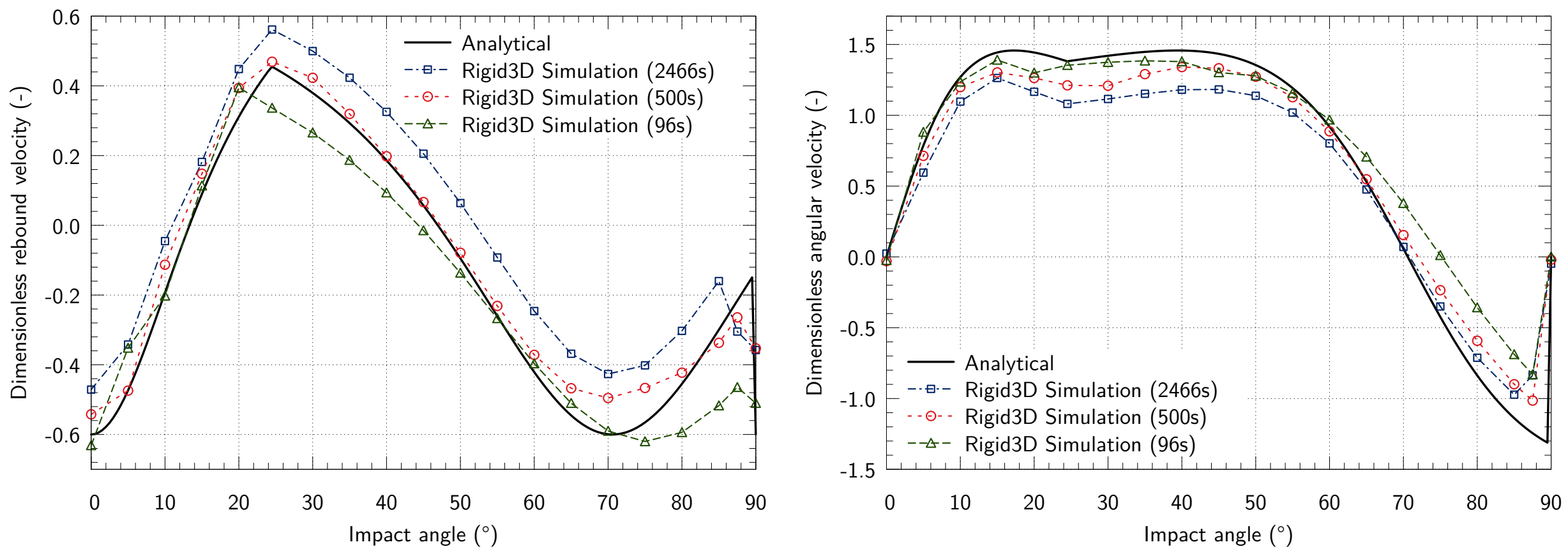

**Figure 12:** Post-impact velocities as a function of impact angle $\theta$. Left: dimensionless translational velocity $V_z^+/V_z^-$; right: dimensionless angular velocity $\omega_y^+ R_b/V_z^-$. The tablet is approximated with 96, 500 and 2466 spheres to represent the coarse, intermediate and fine resolution.

Comparison of the post-impact translational and angular velocities between the simulation results and the analytical solution computed by Eqns. (9-10) is plotted in Figure 12. Overall, the Rigid3D simulation gives relatively good agreement with the analytical solution. Surprisingly, the predicted rebound velocity from the 500-sphere multi-sphere model has the minimum deviation from the analytical solution, except at orientations $\theta \in [65^\circ, 75^\circ]$ ; while the maximal deviation is from the 2466-sphere model where the corner/edge looks sharp enough as shown in Figure 11(f). For the predicted angular velocities at impact angles $\theta \in [0^\circ, 60^\circ]$, the 96-sphere model gives better agreement the finer models, while for $\theta \in (60^\circ, 90^\circ]$ the 2466-sphere model has the minimum deviation.

We can see that the simulation results are not consistent with the increasing resolution of the multi-sphere models, i.e. the sharpness of the tablet's edge in this case. It is because the post-impact dynamics are very sensitive to the geometry at the contact point. For the true tablet, there should be only one contact point in the DEM simulation due to its convex shape, either on the spherical cap or the edge; while for the multi-sphere models with bumpy surface, there is often more than one contact point, as multiple component spheres can be in contact with the flat wall, these contact points also continuously change when the impact angle changes. Therefore randomness occurs in the calculated contact force and moment arm as reflected in Figure 12. Nevertheless, the post-impact dynamics is relatively well predicted by our simulation using the multi-sphere models of different resolutions.

## 5.2 Static packing in containers

To further validate the effectiveness of the proposed algorithm to approximate realistic objects, tow more shapes: M&M candy and the Stanford bunny, are presented to showcase the multi-sphere representations for simple and complex objects, respectively. The geometry of the M&M candy can be approximately described by an oblate spheroid: $\frac{x^2+y^2}{a^2} + \frac{z^2}{c^2} = 1$, where $a = 6.585$ mm, $c = 3.395$ mm; while the Stanford bunny model is often used to test various computer graphics algorithms. The voxel size is set as $d_{eq}/100$, for the object body discretization with surface-conforming grid as demonstrated in Figure 5(e). The computed multi-sphere models for the M&M candy and Stanford bunny are plotted in Figure 13 and Figure 14, respectively. The number of component spheres and their corresponding volume coverage are summarized in Table 2.

**Table 2:** Number of component spheres for increasing resolutions in volume coverage (%)

| | 95.0% | 99.0% | 99.9% | 99.99% |
|---|---|---|---|---|
| M&M candy | 48 | 181 | 461 | 675 |
| Stanford bunny | 255 | 1147 | 2855 | 4090 |

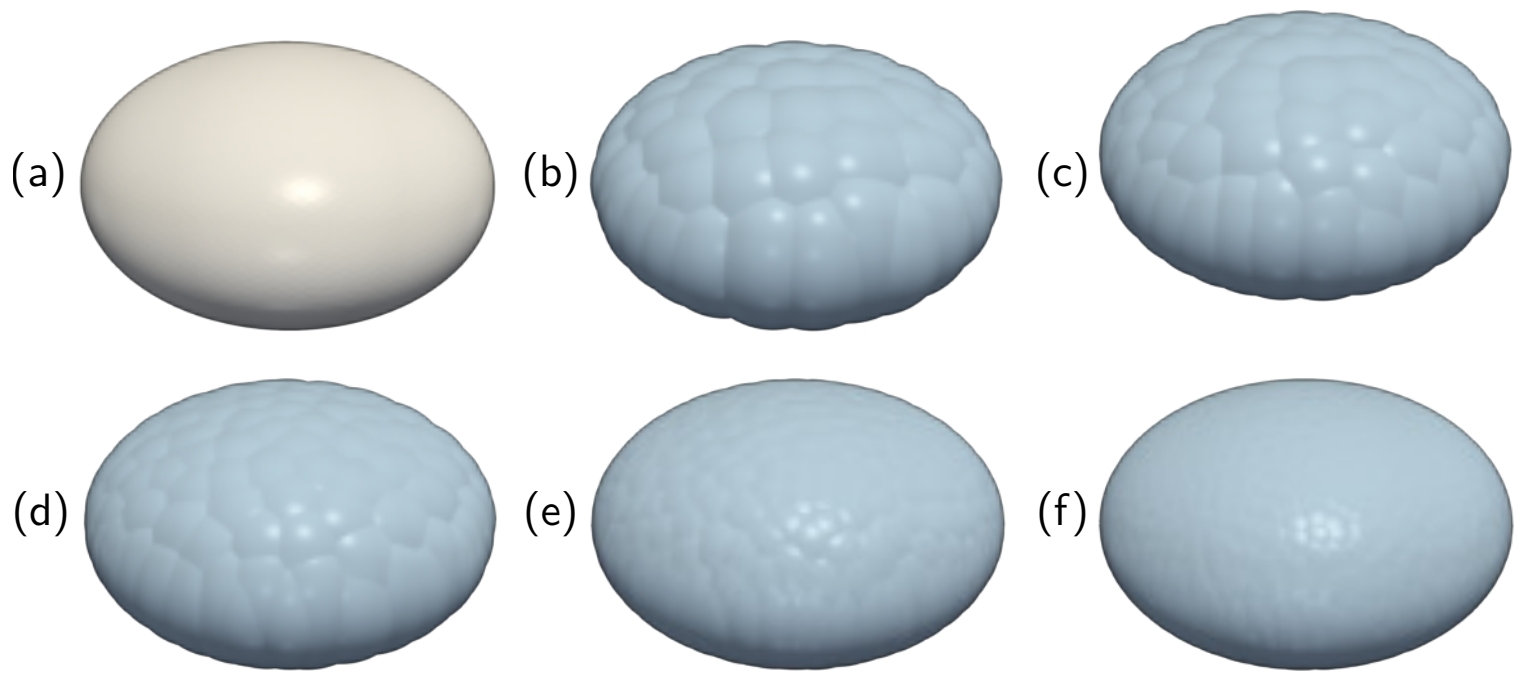

**Figure 13:** The M&M candy multi-sphere models with increasing resolution in number of spheres and volume coverage. (a) original geometry; (b) 60 spheres, 96.02%; (c) 80 spheres, 97.05%; (d) 120 spheres, 98.15%; (e) 265 spheres, 99.50%; (f) 790 spheres, nearly 100% coverage.

Obviously the object shape is the key factor that affects the number of component spheres for the computed multi-sphere models. For rounded objects like ellipsoids, normally only a few dozens of spheres are required for an acceptable level of approximation (e.g. coarse resolution, coverage $\geq$ 95%). If an object has sharp edges/corners, or parts of the object are thin or narrow, a lot small spheres are needed to replicate these geometric features. As shown in Table 2, at volume coverages 95% and 99%, the corresponding component spheres are 48 and 181 for the M&M candy; 255 and 1147 for the Stanford bunny which are about 6 times more than the former. It can be seen in Figure 14(d) that the multi-sphere model of the Stanford bunny is very close to the original shape, except for some small details (e.g. ears) that need a lot of small spheres, e.g. in finer models as shown in Figure 14(e-f). In most cases, multi-sphere models with a resolution of volume coverage 99% for any shape are sufficient to capture the bulk properties of a granular system in DEM simulations.

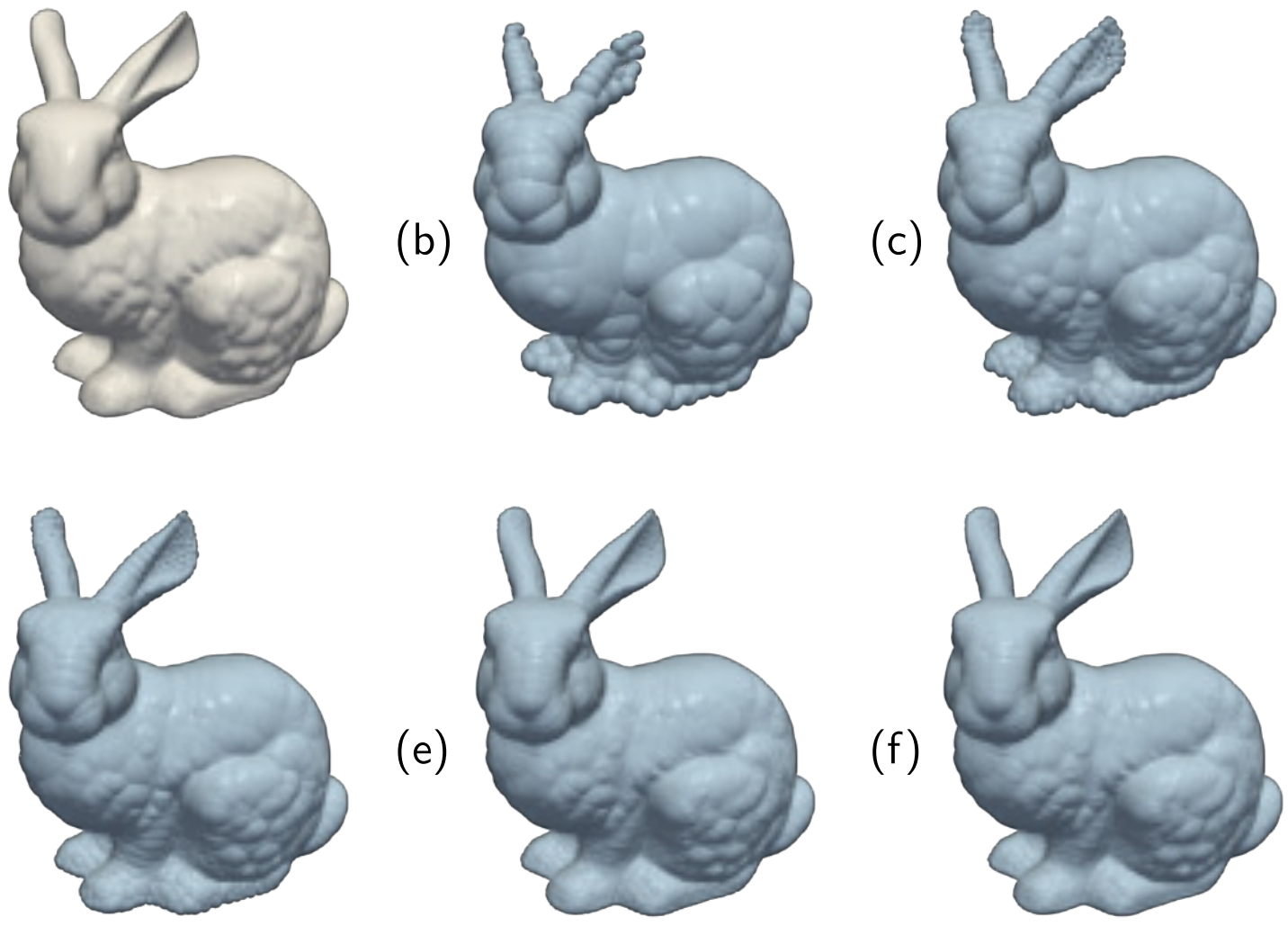

**Figure 14:** The Stanford bunny multi-sphere models with increasing resolution in number of spheres and volume coverage. (a) original mesh model: 69630 triangles, 34817 vertices; (b) 255 spheres, 95.0%; (c) 600 spheres, 97.70%; (d) 1147 spheres, 99.0%; (e) 3000 spheres, 99.92%; (f) 4795 spheres, nearly 100% coverage.

**Table 3:** Parameters used in the M&M candy packing simulation

| Parameter | Value |
|---|---|
| Cylinder diameter (mm) | 50.8 |
| Cylinder height (mm) | 203.2 |
| Number of particles | 250 |
| Density (kg/m$^3$) | 1377.0 |
| Young's modulus (Pa) | $5.0 \times 10^7$ |
| Poisson's ratio | 0.29 |
| Coefficient of friction | $0.3^{(pw)}$, $0.35^{(pp)}$ |
| Coefficient of restitution | 0.5 |
| Time step $\Delta t$ (s) | $5.0 \times 10^{-7}$ |

Superscripts $pw$ and $pp$ denote particle-wall and particle-particle coefficients.

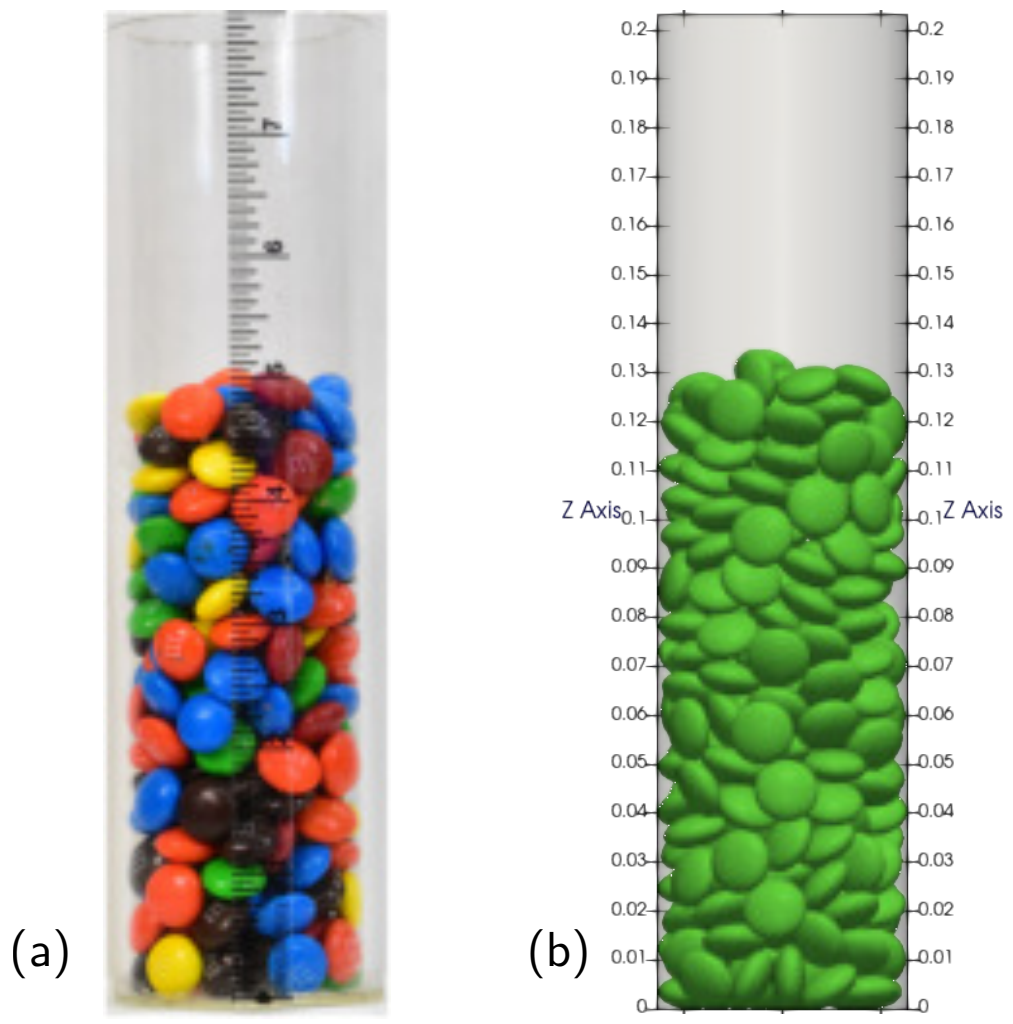

**Figure 15:** Packing of 250 M&M candies in a cylindrical container. (a) experiment carried out by Gao et al. [19], (b) Rigid3D simulation with a resolution of 265 spheres per clump.

To validate the generated multi-sphere models in real applications, two tests were carried out: (1) static packing of 250 M&M candies in a cylindrical container and compared with the experiment reported by Gao et al. [19]; (2) numerical simulation of 10 Stanford bunnies falling in a rectangular container.

In the static packing experiment, 250 particles (i.e., M&M candies) were dropped into the container one by one, while 3~5 particles with random orientation and position are generated in the Rigid3D simulation, and fall under gravity until they settle in the container, because the final packing height still has good agreement with the experimental result. The information about the container geometry, particle's properties and DEM simulation parameters is summarized in Tablet 3. Final states of the static packing from the experiment and the simulation are compared in Figure 15. We can see that the packing height in the Rigid3D simulation has a satisfying agreement with the experimental result. The measured packing heights are 130.0 ± 1.0 mm in the experiment, and 131.2 ± 1.6 mm in the simulation, which is slightly higher than the experimental result. One of the possible reasons is the added particle surface friction due to the clump's bumpy surface as shown in Figure 13(e).

It seems that the Rigid3D simulation is able to closely predict the experimental result. Thus, we could explore more particle and container shapes through the numerical experiments that might be considered as an alternative to the real experiments. Using the same material properties and simulation setup listed in Table 3, we drop 10 rigid Stanford bunnies one by one from the top of a rectangular container as shown in Figure 16. After a few bunnies settling on bottom of the container, a plausible behaviour between rigid bodies, and rigid bodies to wall was observed. The simulation was repeated a couple of times without changing any simulation parameters, while this does not guarantee the same result, because the generated clumps have random initial orientations. With slight disturbance in material properties, initial conditions and un-modeled microstructure of particle's surface, the outcomes may vary significantly, and we may only obtain physically

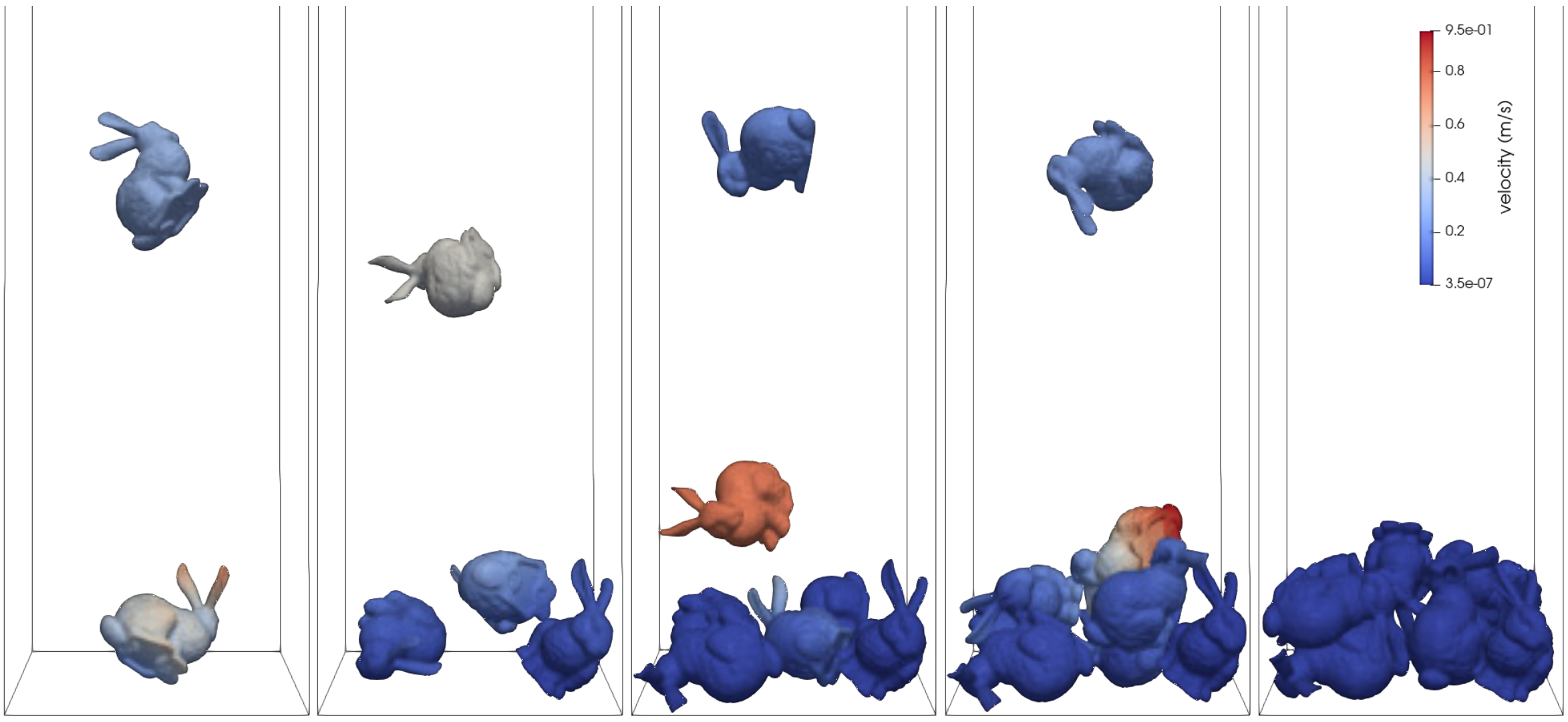

**Figure 16:** Sequential screenshots of 10 Stanford bunnies falling one by one into a rectangular container. The clump resolution here is 3000 component spheres, which corresponds to 99.92% volume coverage.

plausible results [24].

## 5.3 Optimal clump resolution

In order to efficiently simulate a granular system with large number of particles, we want the number of spheres per clump as few as possible, as long as the mechanical behaviour of the system is approximately captured. Therefore, a possible optimal clump resolution (i.e., minimum number of spheres) needs to be found prior to the actual DEM simulations.

Non-spherical particle flow inside a rotating drum is a good example, because of its broad range of industrial applications for processing granular materials. The dynamic angle of repose (AoR) of the particle bed in a rotating drum can be used to find the possible optimal clump resolution. The basic idea is to gradually increase the clump resolution (e.g. from 80% volume coverage) until the averaged dynamic angle of repose converges, which indicates the granular system is dynamically stable with the minimum clump resolution.

Using the same DEM parameters and material properties from the test case of static packing of M&M candies in section 5.2, a numerical experiment was performed for M&M candies tumbling in a cylindrical and rotating drum. The M&M candies are identical to the previous case, and are modeled with increasing resolution as shown in Figure 13. In order to find the possible optimal clump resolution, four resolutions namely 60, 80, 120 and 265 component spheres, which correspond to 96%, 97%, 98% and 99.5% volume coverage respectively, were tested in the numerical simulations. All the clump models are scaled slightly (¡ 100% volume coverage) such that their volumes are identical to the prototype. The drum contained 130 M&M candies, and was tumbled at a rotational speed of 25 rpm. Once the particle bed is more or less stable dynamically in the drum as shown in Figure 17, 10 snapshots were taken every half second for each case with a certain clump resolution, and then an image processing routine presented in [28] is used here to estimate the averaged dynamic angle of repose (DAoR). For each snapshot a free surface profile (e.g. the 2D dotted curve in Figure 17c) can be obtained. Then a linear least squares (LLS) data fitting is used to evaluate the angle of repose. A number of 50∼100 control points evenly distributed on the profile curve may be adequate for the LLS fitting. Note that both the front and the rear profiles are considered, thus the dynamic angle of repose for each snapshot is the averaged value of the two.

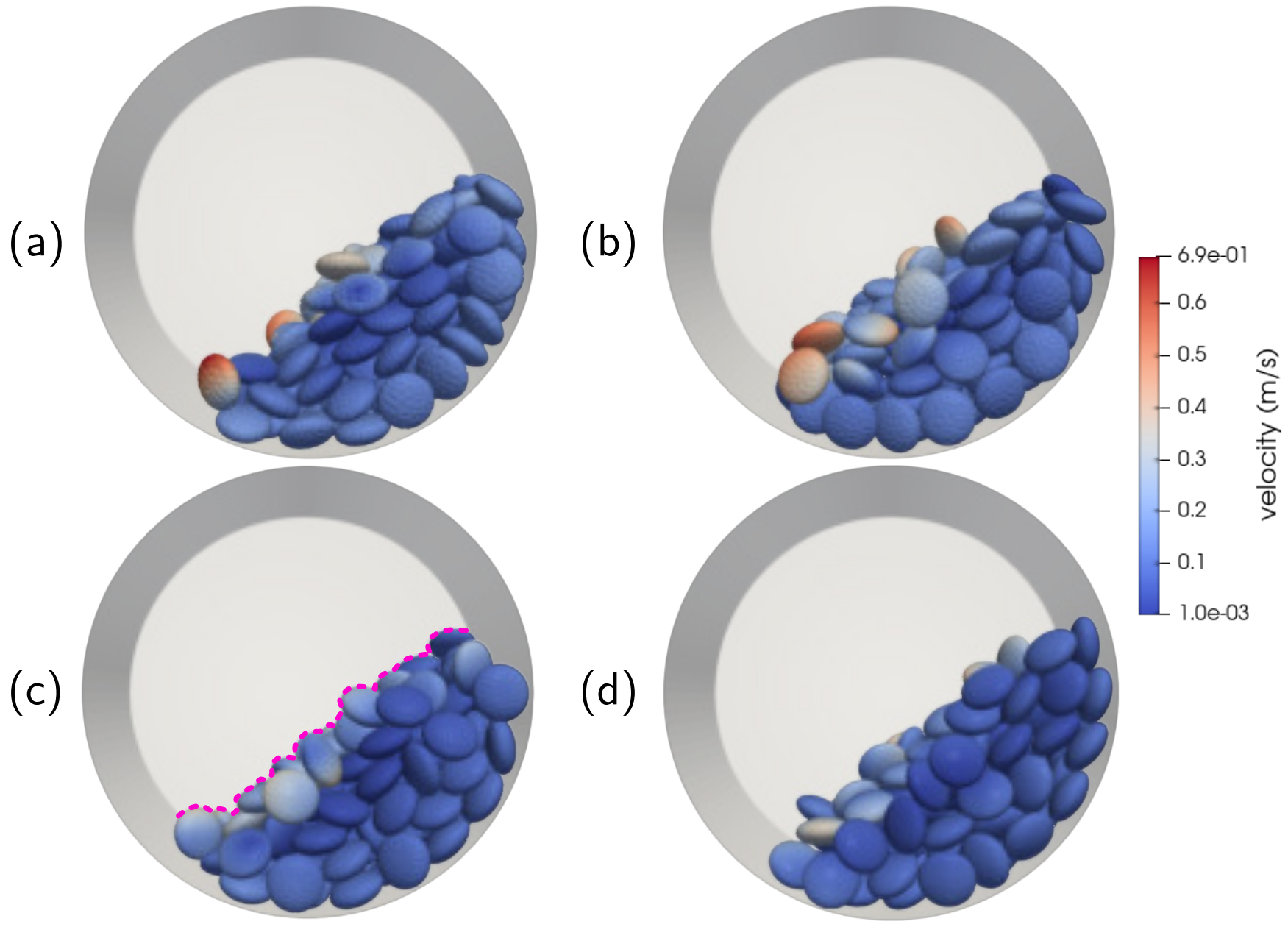

**Figure 17:** Simulation of 130 M&M candies in a rotating drum at a rotational speed 25 rpm, with increasing particle resolution in number of component spheres. (a) 60 spheres (96%); (b) 80 spheres (97%); (a) 120 spheres (98%); (a) 265 spheres (99.5%);

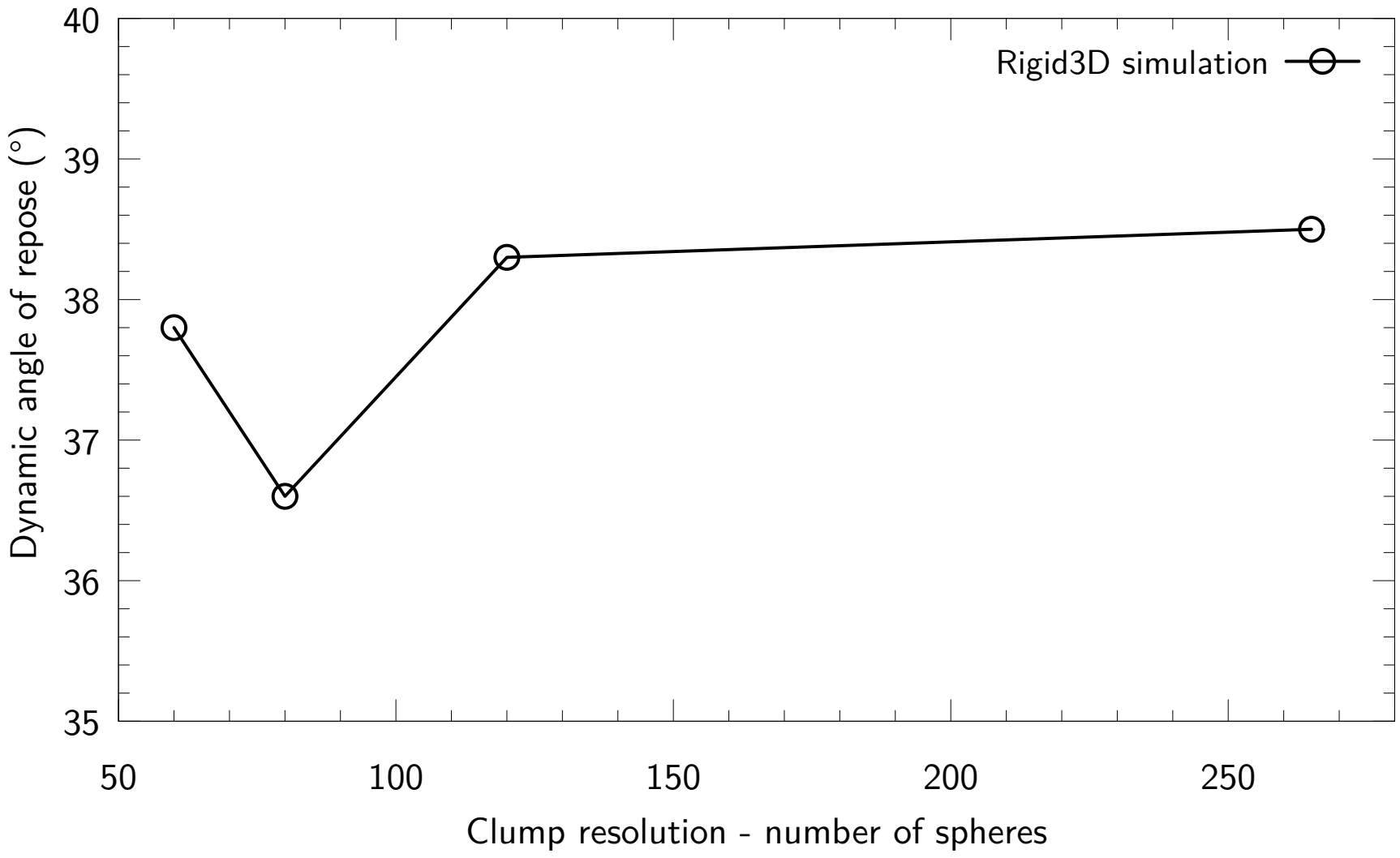

**Figure 18:** Averaged dynamic angle of reposes of the particle beds with increasing clump resolution: 60, 80, 120 and 265 spheres.

The averaged dynamic angle of reposes for each clump resolution is plotted in Figure 18. The results indicate that an increasing clump resolution leads to an increase of the dynamic angle of repose. This can be explained by the fact that more spheres per clump will increases the surface roughness, thus the increased interlocking hinders particles to move freely down the surface of the particle bed, and tends to form a steeper slope [28, 54]. Nevertheless, if a particle is approximated with a relatively coarse resolution, the dynamic angle of repose might be larger than that of finer resolution. In our case, the DAoR at the resolution of 60 spheres is c.a. 37.8°, which is larger that the DAoR - 36.6°at the resolution of 80 spheres. This is because a very coarse approximation has rather large concave corners on the particle surface, thus increases the interlocking between particles. In this case, it is not recommended to use the coarse version for DEM simulations, even though the dynamic angle of repose is close to the converged value, as it does not reflect the real shape characteristics in the particle flow, thus it might lead to large deviation on other types of particle flow such as hopper discharge and mixing, etc. The difference in DAoR (38.3°) at the resolutions of 120 and 265 spheres is marginal: 38.3°vs 38.5°,taking into account the randomness in taking snapshots. Even though there is no physical experiment to compare with, we could still consider the measured DAoR at the high clump resolution (265 component spheres, 99.5% volume coverage) as a reference value. Therefore in this case c.a. 120 spheres per clump (98% volume coverage) might be the optimal resolution for granular flow in a rotating drum.

## 6. Conclusion

A novel multi-sphere approach with a combined 3D thinning and greedy set-covering algorithm has been proposed to approximate realistic particles. Compared to previous work either based on the particle's medial-surface or discretized particle body with uniform grid, the proposed approach combines the best of both worlds in terms of approximation accuracy and computational efficiency, taking the advantages of both without the drawbacks described in the introduction section. Most importantly, for a given approximation accuracy in volume coverage, generated clumps using the proposed approach in this work have the least number of spheres among other approaches, thanks to the use of voxelized medial-surface and greedy set-covering algorithms.

Several object shapes including bi-convex tablet, M&M candy and Stanford bunny were approximated by the proposed algorithm with increasing resolutions. These multi-sphere models were then used in the numerical tests carried out by our newly developed DEM code Rigid3D, and the results were compared with either analytical or experimental data. The tablet-wall impact case showed that acceptable results were predicted, however not always consistent with the clump resolutions, because of the random roughness on the surface and sharp corners. Nevertheless, the effect of randomness may cancel each other for a granular system of many particles, given the fact that the predicted fill height of 250 M&M candies Rigid3D packing in a cylindrical container by the DEM simulation has an excellent agreement with the experimental result. The proposed algorithm is capable of modeling very complex shapes with around one thousand spheres with very high resolution in terms of volume coverage. For example, the Stanford bunny can be well approximated by only 1147 spheres with 99% volume coverage. A very satisfying result can be observed from the numerical test of 10 bunnies falling in the rectangular container.

The key to efficiently simulate a large granular system is to keep the clump resolution as low as possible, as long as the mechanical behaviour of the system can be captured quantitatively. In order to find an optimal clump resolution, the particle flow of 130 M&M candies in a rotating drum was investigated numerically with increasing clump resolution. The dynamic angle of repose starts to converge at the resolution of 98% volume coverage (120 component spheres), which implies a possible optimal resolution is found for this type of particle flow. It seems that the clump volume coverage is an appropriate indicator to quantify the particle's approximation accuracy with the multi-sphere models.